\documentclass{iopart}

\usepackage{graphicx,color}

\expandafter\let\csname equation*\endcsname\relax
  \expandafter\let\csname endequation*\endcsname\relax
\usepackage{amsmath}
\usepackage{amssymb}
\usepackage{amsfonts}
\usepackage{bm}
\usepackage[T1]{fontenc}
\usepackage{iopams}
\newcommand{\vect}[1]{\bm{#1}}
\newcommand{\rr}{\vect{r}}
\newcommand{\tens}[1]{\mbox{\textbf{\textit{\textsf{#1}}}}}

\newcommand{\trace}{\operatorname{tr}}
\newcommand{\trans}{\mathsf{T}}
\newcommand{\dif}{\mathrm{d}}
\newcommand{\mi}{\mathrm{i}}
\newcommand{\me}{\mathrm{e}}
\newcommand{\im}{\mathrm{Im}\,}

\newcommand{\dela}{{\nabla}_{\!\rma}}
\newcommand{\delb}{{\nabla}_{\!\rmb}}
\newcommand{\leda}{\overleftarrow{\nabla}_{\!\rma}}
\newcommand{\ledb}{\overleftarrow{\nabla}_{\!\rmb}}
\newcommand{\rma}{\mathrm{A}}
\newcommand{\rmb}{\mathrm{B}}
\usepackage{cite}
\usepackage{soul}

\begin{document}

\setlength{\parindent}{0cm}

\title{Medium-assisted van der Waals dispersion 
interactions involving chiral molecules}

\author{Hassan Safari}
\address{Department of Photonics, Graduate University of Advanced Technology, Kerman, Iran}

\author{Pablo Barcellona}
\address{Physikalisches Institut, Albert-Ludwigs-Universit\"at
	Freiburg, Hermann-Herder-Str. 3, 79104 Freiburg, Germany}

\author{Stefan Yoshi Buhmann}
\address{Physikalisches Institut, Albert-Ludwigs-Universit\"at
Freiburg, Hermann-Herder-Str. 3, 79104 Freiburg, Germany}
\address{Freiburg Institute for Advanced Studies,
Albert-Ludwigs-Universit\"at Freiburg, Albertstr. 19, 79104 Freiburg,
Germany} 

\author{A.~Salam}
\address{Department of Chemistry, Wake Forest University, Winston-Salem, North Carolina 27109, USA}
\address{Physikalisches Institut, Albert-Ludwigs-Universit\"at
Freiburg, Hermann-Herder-Str. 3, 79104 Freiburg, Germany}
\address{Freiburg Institute for Advanced Studies,
Albert-Ludwigs-Universit\"at Freiburg, Albertstr. 19, 79104 Freiburg,
Germany}

\begin{abstract}
The van der Waals dispersion interaction between two chiral molecules
in the presence of arbirary magnetoelectric media is derived using perturbation theory. To be general, the molecular polarisabilities are assumed to be of electric, paramagnetic and diamagnetic natures and the material environment is considered to possess a chiral electromagnetic response. 
The derived formulas of electric--chiral, paramagnetic--chiral, diamagnetic--chiral and chiral--chiral interaction potentials  when added to the previously obtained 
contributions in literature, form a complete set of dispersion interaction formulas.
We present them in a unified form making use of 
electric--magnetic duality. 
As an application, the case of two anisotropic molecules in free space is considered where we drive the retarded and non-retarded limits with respect to intermolecular distance.

\end{abstract}
\submitto{\NJP}

\maketitle

\section{Introduction}
\label{intro}
The non-superposability of an object on its mirror image classifies it as 
chiral. A familiar example is provided by left and right hands. Molecules that 
are chiral lack an improper axis of rotation, exist as 
enantiomeric pairs, and exhibit optical activity \cite{atkins1968,power1971,
caldwell1971, charney1985,Barron2009}, that is, they are able to rotate the 
plane of polarization of light either to the left or to the right, and are so 
termed laevorotatory or dextrorotatory. Other manifestations of molecular 
handedness include differential absorption (circular dichroism) 
\cite{Power1974} 
and differential scattering (Rayleigh and Raman) 
\cite{Barron1971} 
of circularly polarized light, their nonlinear analogues 
\cite{fischer2005}, 
as well as other chiroptical spectroscopies that depend quadratically or on higher powers of the strengths of electromagnetic fields, such as sum-frequency and second-harmonic generation 
\cite{giordmaine1965, fischer2010, andrews2018-1}. 
Many of these phenomena have also been predicted to occur when the incident radiation is of the structured type \cite{andrews2004,cameron2014,cameron2017,forbes2018,babiker2018,forbes2019-1,
forbes2019-2, ForbesUnpublished}.

Because chiral compounds possess reduced or no elements of symmetry, selection rules normally in operation that are used to determine whether spectroscopic transitions are allowed or forbidden are considerably relaxed, or no longer apply. This enables multipole moments such as the magnetic dipole and electric quadrupole to feature to leading order in chiroptical processes, in addition to the usually dominant electric dipole moment. Accounting from the outset for both the charge and current density distributions of a collection of protons and electrons that form atoms and molecules is therefore necessary. Furthermore, Maxwell's equations are needed to describe intrinsic electromagnetic effects due to the sources, as well as those arising from any applied radiation fields. Due to the microscopic nature of such elementary charged particles, both radiation and matter ought to be treated rigorously subject to the laws of quantum mechanics \cite{dirac1958}. Hence the theory of molecular QED \cite{craig2012,salam2010,andrews2018-2} naturally lends itself as the obvious means by which to investigate fundamentally the interactions between photons and electrons, and any effects due to the handedness of molecules.

Non-relativistic QED theory in the Coulomb gauge has not only been successfully applied to a whole host of linear and nonlinear optical processes, but also permits inter-particle interactions to be studied using the same formalism. Well-known examples include resonant transfer of electronic excitation energy \cite{craig2012, salam2010, craig1992, daniels2003, grinter2016, salam2018}, and dispersion forces between two and three atoms or molecules 
\cite{craig2012, salam2010, casimir1948,
aub1960, power1993, passante1999, salam2016, passante2018}. For interactions between enantiomers, be they chemically identical or distinct species, both energy transfer and the dispersion energy shift are discriminatory and depend upon the handedness of the molecules 
\cite{mavroyannis1962,craig1971,craig1998-2,jenkins1994-1,
jenkins1994-2, salam1996, salam2000, salam2005}. According to QED theory, the interaction between particles is mediated by the exchange of 
virtual photons -- by definition unobservable but permitted by the time-energy uncertainty principle \cite{andrews2014,salam2015}. In the case of migration of energy, a single virtual photon is responsible for conveying energy from an excited donor moiety to an unexcited acceptor entity. This compares with the dispersion force, which is understood to arise from the exchange of two virtual photons.

In the presence of an environment, the mediating photons are no longer free-space light quanta associated with the vacuum electromagnetic field but become modified by the medium. One approach for dealing with this is to affect a scheme in which the radiation field and the surrounding environment are quantized, resulting instead in medium-assisted photons. This formed the basis for the construction of a macroscopic QED theory in which a magnetoelectric medium is accounted for and described by frequency dependent electric permittivity and magnetic permeability functions \cite{matloob1995,scheel2008,
buhmann2013-1,buhmann2013-2,buhmann2012}. As in molecular QED, its macroscopic counterpart has been cast in terms of minimal- and multipolar-coupling frameworks; it has been applied with great advantage to the calculation of Casimir, Casimir--Polder, and van der Waals forces. For the last type of interaction, this has specifically included the evaluation of pair dispersion potentials in a magnetoelectric medium involving atoms or molecules that are either electrically polarizable or paramagnetically and diamagnetically susceptible, and their behavior has been examined as a function of interparticle separation distance in the non-retarded and retarded coupling regimes \cite{safari2006, buhmann2013-3}.

Within the context of interactions involving optically active species, up to 
now macroscopic QED has only been employed to study 
Casimir--Polder forces 
\cite{Butcher2012, Barcellona2016}, 
i.e.~between a microscopic particle and a macroscopic object, and the van der 
Waals dispersion energy between an electrically polarizable molecule and a 
chiral molecule. While it is known that the last mentioned potential vanishes 
in free space for an isotropic pair of interacting molecules, it was shown 
recently that a non-zero energy shift arises by specifically choosing the 
environment to be a chiral plate 
\cite{Barcellona2017}.
 Not only is the energy finite, but its magnitude and sign may even be controlled and its behavior studied for particular configurations of the chiral molecules and the achiral molecule relative to the chiral plate. For instance, there is maximal enhancement of the interaction when the two particles are aligned parallel to the plate but are each separated a large distance from it. This offers a realistic possibility for the separation of enantiomers, complementing a recent proposal exploiting parity violation in the Casimir--Polder potential when a beam of chiral molecules passes through a pair of chiral mirrors \cite{suzuki2019}.
 
  In the present work, we develop a general and complete macroscopic QED theory for the van der Waals dispersion interaction between two molecules 
  with arbitrary linear dipolar response
  that may or may not be chiral and are described by their electric polarisability, para- and diamagnetic susceptibility, and mixed electric--magnetic dipole (chiral) susceptibility in a magnetoelectric medium. Fourth-order diagrammatic perturbation theory is employed to compute the interaction energy, and explicit contributions involving one or two chiral species are extracted. On inserting the free-space form of the Green's tensor, it is found that the results obtained reduce to those previously derived using molecular QED \cite{salam2010}. 
 
 The paper is organized as follows: The basic formalism for QED
in the presence of chiral magnetoelectric media together with the introduction of molecule-field interaction Hamiltonians are given in Sec.~\ref{formalism}. In Sec.~\ref{potential}
the derivation of electric--chiral, paramagnetic--chiral, and chiral--chiral
van der Waals dispersion potentials are presented via fourth order perturbation theory, followed by diamagnetic--chiral interaction for which the 
third order perturbation theory is 
applied. A unified form of the formulas of various dispersion interaction potentials is derived in
 Sec.~\ref{duality} using the electric--magnetic duality. 
As an application of the obtained formulas, the interaction between two anisotropic molecules in free space is computed in Sec.~\ref{freespace}
where also the retarded and non-retarded limits of intermolecular distances are 
considered. A summary and concluding remarks are provided in Sec.~\ref{conclusion}.

\section{Basic Formalism}
\label{formalism}
The Hamiltonian of a system comprised of two molecules $A$ and $B$ in the presence of a quantized electromagnetic field is given as
\begin{equation}
\label{Ham}
\hat{H} = \hat{H}_F + \sum_{A'=\rma,\rmb}\hat{H}_{A'} +  \sum_{A'=\rma,\rmb}\hat{H}_{A'F},
\end{equation}
where $\hat{H}_F$, $\hat{H}_{A'}$, and $\hat{H}_{A'F}$ denote, respectively, field, 
molecule, and molecule--field interaction Hamiltonians. 
In terms of fundamental photonic annihilation and creation operators, $\hat{\vect{f}}_\lambda$ and $\hat{\vect{f}}_\lambda^\dagger$, the field Hamiltonian reads
\begin{equation}
\label{HF}
\hat{H}_F=\sum_{\lambda=e,m}\int\dif^3 r\int_0^\infty \dif\omega \hbar\omega\hat{\vect{f}}_\lambda^\dagger(\vect{r},\omega)\cdot
\hat{\vect{f}}_\lambda(\vect{r},\omega)
\end{equation}
with $\lambda$ referring to the electric $(\lambda=e)$ or magnetic $(\lambda=m)$ 
nature of the noise source. These operators obey the following commutation relations:
\begin{eqnarray}
&&\label{comm1}
\left[\hat{\vect{f}}_\lambda(\vect{r},\omega),\hat{\vect{f}}_{\lambda'}(\vect{r}',\omega')\right]
=\left[\hat{\vect{f}}_\lambda^\dagger(\vect{r},\omega)
,\hat{\vect{f}}_{\lambda'}^\dagger(\vect{r}',\omega')\right]=\bm{0},\\
&&
\label{comm2}\left[\hat{\vect{f}}_\lambda(\vect{r},\omega),\hat{\vect{f}}_{\lambda'}^\dagger(\vect{r}',\omega')\right]
=\tens{I}\delta_{\lambda\lambda'}\delta(\vect{r}-\vect{r}')
\delta(\omega-\omega'),
\end{eqnarray}  
with $\tens{I}$ being the identity matrix. The ground-state of the field $|\{0\}\rangle$ is defined by
\begin{equation}
\hat{\vect{f}}_\lambda(\vect{r},\omega)|\{0\}\rangle=0,\quad \forall\, \vect{r},\omega,\lambda
\end{equation}
and the excited photonic states are defined by repeated application of the creation 
operator to the ground-state. For example, for single- and two-photon excited states, 
\begin{eqnarray}
&&\label{eq6}
|\bm{1}_\lambda(\vect{r},\omega)\rangle  = \hat{\vect{f}}_\lambda^\dagger(\vect{r},\omega)|\{0\}\rangle,\\
&&\label{eq7}|\bm{1}_\lambda(\vect{r},\omega),\bm{1}_{\lambda'}(\vect{r}',\omega')\rangle =
\frac{1}{\sqrt{2}}\hat{\vect{f}}_\lambda^\dagger(\vect{r},\omega)\hat{\vect{f}}_{\lambda'}^\dagger(\vect{r}',\omega')|\{0\}\rangle.
\end{eqnarray}
The molecular Hamiltonian may be written in terms of unperturbed molecular eigenenergies and eigenstates as
\begin{equation}
\label{HA}
\hat{H}_{A'} =\sum_k E_{A'}^k| k_{A'}\rangle\langle k_{A'}|,
\quad A'=\rma,\rmb.
\end{equation}
In the multipolar coupling scheme, the molecule--field interaction Hamiltonian takes the form \cite{buhmann2013-3}
\begin{equation}
\label{HA'}
\hat{H}_{A'F} = -\hat{\vect{d}}_{A'}\cdot\hat{\vect{E}}(\vect{r}_{A'})
-\hat{\vect{m}}_{A'}\cdot\hat{\vect{B}}(\vect{r}_{A'})
-\frac{1}{2}\hat{\vect{B}}(\vect{r}_{A'})
\cdot\hat{\bm{\beta}}^D_{A'}\cdot\hat{\vect{B}}(\vect{r}_{A'}),
\end{equation} 
with $\hat{\vect{d}}_{A'}$ and $\hat{\vect{m}}_{A'}$ being, respectively, electric and magnetic dipole moment operators of the molecule $A'$ and $\hat{\vect{\beta}}^D_{A'}$ is its operator-valued diamagnetisability tensor 
\begin{equation}
\label{beta}
\hat{\bm{\beta}}^D_{A'} = -\sum_{\alpha\in A'}\frac{q_\alpha^2}{4m_\alpha}(\hat{\bar{r}}_\alpha^2\bm{I}-\hat{\bar{\rr}}_\alpha
\hat{\bar{\rr}}_\alpha)\,,
\end{equation} 
where $q_\alpha$ and $m_\alpha$ are, respectively,
the electric charge and mass of the particle $\alpha$, and $\hat{\bar{\rr}}_\alpha$ is its position vector relative to the center of mass of the molecule.
The electric and magnetic fields can be expressed as linear combinations of the fundamental photonic operators.
To this end, we introduce 
$\bm{\mathcal{R}}_{\lambda\lambda'}(\rr,\omega)$ ($\lambda,\lambda' = e,m$) as the $3\times 3$
blocks of the $6\times 6$ tensor $\bm{\mathcal{R}}(\rr,\omega)$,
\begin{equation}
\bm{\mathcal{R}} = 
\left(
\begin{matrix}
\bm{\mathcal{R}}_{ee} &
 \bm{\mathcal{R}}_{em}\\
\bm{\mathcal{R}}_{me} & 
\bm{\mathcal{R}}_{mm}
\end{matrix}
\right),
\end{equation}
which is the response tensor of the magnetoelectric material environment \cite{Butcher2012} defined, for locally responding media, according to
\begin{equation}
\label{R}
\bm{\mathcal{R}}\cdot \bm{\mathcal{R}}^\top =
\left(
\begin{matrix}
\varepsilon_0\im(\bm{\varepsilon}-
\bm{\kappa}^\top\cdot\bm{\mu}^{-1}\cdot\bm{\kappa}) &
 -\mi\sqrt{\frac{\varepsilon_0}{\mu_0}}\im(\bm{\kappa}^\top\cdot\bm{\mu}^{-1})\\
\mi\sqrt{\frac{\varepsilon_0}{\mu_0}}\im(\bm{\mu}^{-1}\cdot\bm{\kappa}) & 
-\frac{1}{\mu_0}\im \bm{\mu}^{-1}
\end{matrix}
\right).
\end{equation}
In Eq.~\eqref{R} $\bm{\varepsilon}=\bm{\varepsilon}(\rr,\omega)$ and $\bm{\mu}=\bm{\mu}(\rr,\omega)$ are, respectively, the relative electric permittivity and magnetic 
permeability tensors of the media, and 
$\bm{\kappa}=\bm{\kappa}(\rr,\omega)$ is the chirality tensor responsible for the contribution of an 
electric effect to a magnetic response and vise versa. The electric and magnetic fields are given as
\begin{eqnarray}
&&\label{E}
\hat{\vect{E}}(\vect{r})=\sum_{\lambda=e,m}\int\dif^3 r'\int_0^\infty \dif\omega \tens{G}_\lambda(\vect{r},\vect{r}',\omega)\cdot\hat{\vect{f}}_\lambda(\vect{r}',\omega)+ H.c.,\\
&&\label{B}
\hat{\vect{B}}(\vect{r})=\sum_{\lambda=e,m}\int\dif^3 r'\int_0^\infty \frac{\dif\omega}{\mi\omega}\nabla\times 
\tens{G}_\lambda(\vect{r},\vect{r}',\omega)\cdot
\hat{\vect{f}}_\lambda(\vect{r}',\omega)+ H.c.\,,
\end{eqnarray}
where $\tens{G}_{e}$ and $\tens{G}_m$ are the mode tensors introduced in terms of the Green tensor $\tens{G}$,
\begin{multline}
\label{Ge}
	\tens{G}_\lambda(\vect{r},\vect{r}',\omega)\\
=
	-\mi\mu_0\omega\sqrt{\frac{\hbar}{\pi}}
	\left[\mi\omega\tens{G}(\vect{r},\vect{r}',\omega)\cdot
	\bm{\mathcal{R}}_{\lambda e}(\vect{r}',\omega)
	+
	\tens{G}(\vect{r},\vect{r}',\omega)\times\overleftarrow{\nabla}'\cdot
	\bm{\mathcal{R}}_{\lambda m}(\vect{r}',\omega)\right],
\end{multline}
with $\overleftarrow{\nabla}'$ denoting differentiation from the left.
The Green tensor 
$\tens{G}(\vect{r},\vect{r}',\omega)$ obeys the differential equation
\begin{eqnarray}
&\hspace{-.9in}\nabla\times\bm{\mu}^{-1}\cdot\nabla\times\tens{G}
+\frac{\omega}{c}\nabla\times\bm{\mu}^{-1}\cdot\bm{\kappa}
\cdot\tens{G}
+\frac{\omega}{c}\bm{\kappa}^\top\cdot\bm{\mu}^{-1}\cdot
\nabla\times\tens{G}
\nonumber\\
&\hspace{.6in}-\frac{\omega^2}{c^2}\left(\bm{\varepsilon}
-\bm{\kappa}^\top\cdot\bm{\mu}^{-1}\cdot\bm{\kappa}\right)\cdot\tens{G}
=\tens{I}\delta(\vect{r}-\vect{r}'),
\end{eqnarray}
($\textit{\textsf T}^\top_{ij}
={\textit{\textsf T}}_{ji}$). All geometric and magnetoelectric properties of the environment are taken into account via $\bm{\varepsilon}(\vect{r},\omega)$, $\bm{\mu}(\vect{r},\omega)$, and $\bm{\kappa}(\vect{r},\omega)$. Furthermore, the Green tensor obeys the Schwarz reflection principle
\begin{equation}
\label{Schwartz}
\tens{G}(\vect{r},\vect{r}',-\omega)
=\tens{G}^\ast(\vect{r},\vect{r}',\omega^\ast),
\end{equation}
Onsager reciprocity
\begin{equation}
\label{Onsager}
\tens{G}(\vect{r},\vect{r}',\omega)
=\tens{G}^\top(\vect{r}',\vect{r},\omega),
\end{equation}
and a useful integral relation  \cite{dung2003, buhmann2004}
\begin{equation}
\label{intrel}
\sum_{\lambda=e,m}\int\dif^3 s \tens{G}_\lambda(\vect{r},\vect{s},\omega)\cdot
\tens{G}_\lambda^{\ast\top}(\vect{r}',\vect{s},\omega) =
\frac{\hbar\mu_0\omega^2}{\pi}\,\im \tens{G}(\vect{r},\vect{r}',\omega).
\end{equation}
These properties will be used in the evaluation of matrix elements appearing in perturbation
formulas.
\section{Interaction Potential}
\label{potential}

For ground-state molecules the van der Waals dispersion interaction is mediated by two virtual-photon exchanges between the molecules accompanied by internal transitions.  In achiral molecules the molecular eigenstates are also the eigenstates of the parity operator. Hence, the resulting expression for the interaction potential is obtained such that every molecule may be considered as a superposition of an electric, a paramagnetic, and a diamagnetic contribution for which the molecular transitions are, respectively, electric, paramagnetic, or diamagnetic type only (see Ref.~\cite{buhmann2013-3}).
For chiral molecules, however, each molecular transition may possess interdependent electric and magnetic dipole moments and additional higher multipole moments. This leads to an additional chiral contribution to the interaction potential, to which the rest of this section is devoted. 

In order to calculate the interaction energy between two ground-state 
molecules $\rma$ and $\rmb$ in the presence of a medium-assisted electromagnetic field we 
use perturbation theory. To do so, we consider the sum of the
field and molecular Hamiltonians as unperturbed Hamiltonian, and the sum of the molecule--field
interaction Hamiltonians,
$\hat{H}_{\mathrm{AF}}
\!+\!\hat{H}_{\mathrm{BF}}\equiv\hat{H}_{\rm{int}}$, as perturbation. The electric or paramagnetic transitions in the molecules depend linearly on the electric and magnetic fields, respectively, and are
associated with the absorption or emission of a single virtual photon, while the diamagnetic coupling involves a two-photon transition. Hence, in the calculation of the energy shift, different orders of perturbation theory are required for diamagnetic molecules in comparison to electric and paramagnetic ones.

\subsection{Electric--chiral, paramagnetic--chiral, and  chiral--chiral interaction}
\begin{figure}[!ht]
	\hspace{.3in}\includegraphics[width=.9\linewidth]{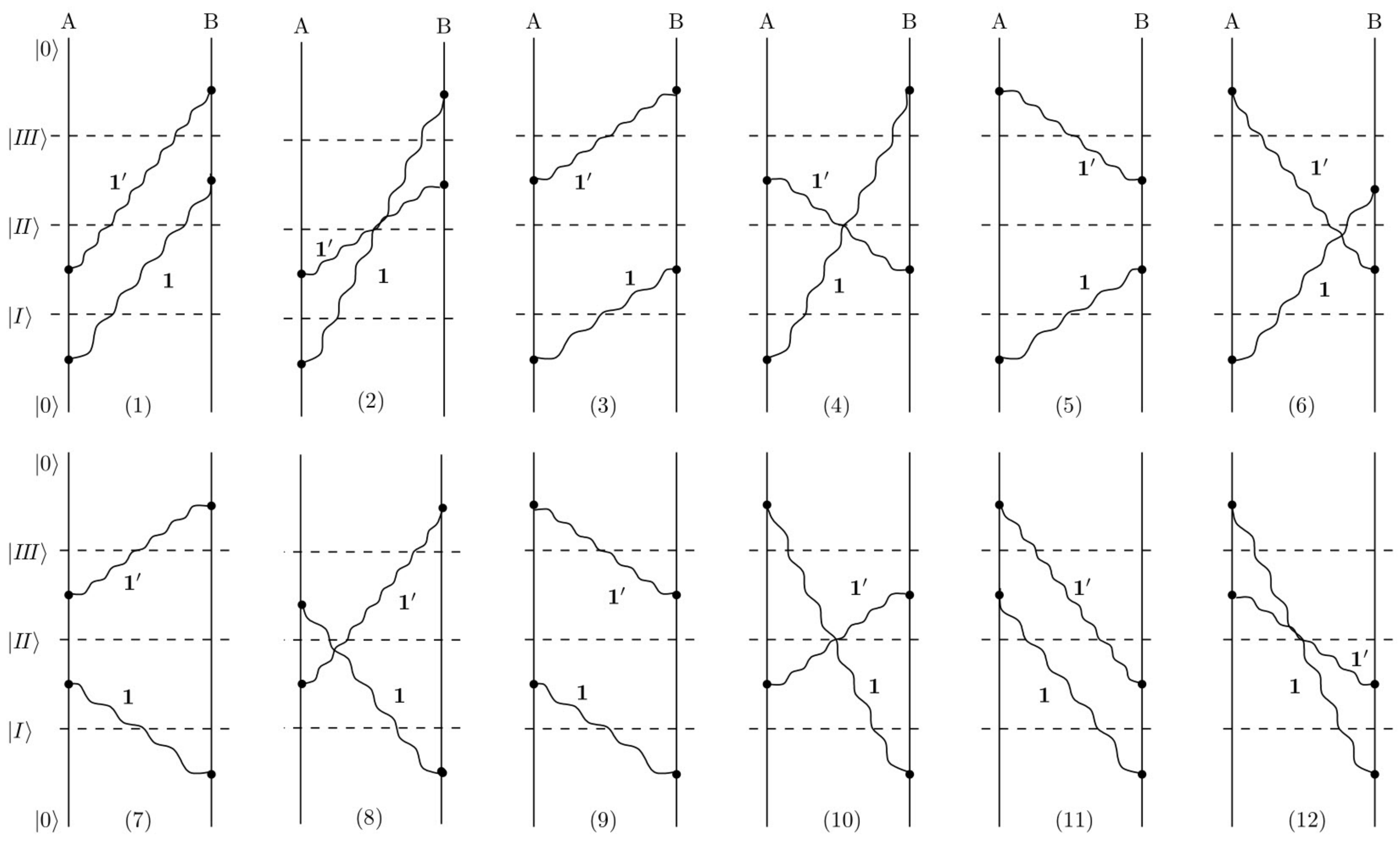}
\caption{\label{fig1}
Two-virtual photon exchange diagrams contributing to the van der Waals
dispersion potential. The various intermediate states required in Eq.~(\ref{p4}) are readily obtained.} 
\end{figure}

The lowest order perturbation leading to the interaction 
potential,
involving an electric, paramagnetic, or a chiral molecule
that interacts with a second species 
which
 is chiral via two virtual photon exchange is the fourth order \cite{buhmann2013-3, safari2006},
\begin{equation}
\label{p4}
U(\rr_{\rma},\rr_{\rmb})=
-\sum_{I,I\!I,I\!I\!I\neq0}
\hspace{-1ex}
\frac{\langle 0|\hat{H}_{\rm{int}}
	|I\!I\!I\rangle
	\langle I\!I\!I|\hat{H}_{\rm{int}}|I\!I\rangle\langle I\!I|\hat{H}_{\rm{int}}|I\rangle
	\langle I|\hat{H}_{\rm{int}}|0\rangle}
{(E_{I\!I\!I}-E_0)(E_{I\!I}-E_0)(E_{I}-E_0)}\,,
\end{equation}
with the ground state of the total system denoted by $|0\rangle\equiv|0_\rma\rangle|0_\rmb\rangle|\{0\}\rangle$.
The numerator is comprised of multiplication of four matrix elements, 
each corresponding to an event in which one of the molecules undergoes a transition 
accompanied by emission or absorption of a virtual photon. The photon emitted by one of the 
molecules has to be absorbed by the other to contribute to the coupling between molecules.  
In order to take into account a complete set of intermediate states in calculating the
right hand side of Eq.~(\ref{p4}), one may use a diagrammatic language depending on the time-ordered sequence of the propagation of two virtual photons, as sketched in Fig.~\ref{fig1}. 
In every single diagram shown, each molecule undergoes two transitions,
from the ground state to an excited state and back. 

The chirality contribution of a molecule to the interaction potential comes from processes in which an electric upward transition in the molecule is followed by a magnetic downward one in the same molecule, or vice versa. 
To this order of multipole approximation it can be decomposed into three parts: electric--chiral, paramagnetic--chiral, 
and chiral--chiral, each of which is examined below.	            

\begin{figure}[!ht]
	\centering
	\includegraphics[width=.4\linewidth]{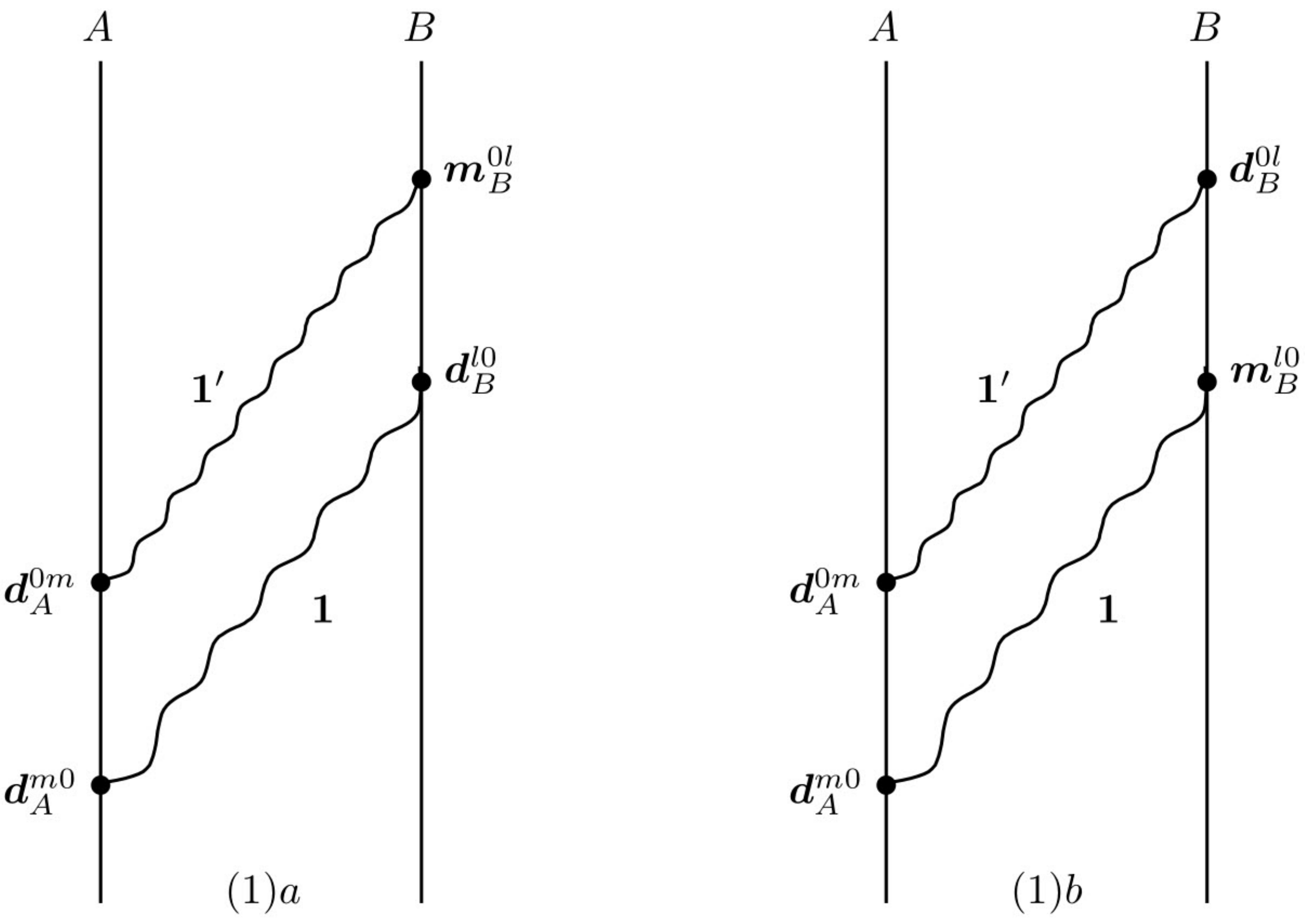}
	\caption{\label{fig2} Two processes corresponding to diagram $(\mathrm{1})$ in Fig.~\ref{fig1}, for electric--chiral interaction.
	}
\end{figure}

\paragraph{Electric--chiral interaction.}
Let us consider an electric molecule $\rma$ and a chiral molecule $\rmb$ located 
respectively at positions $\rr_\rma$ and $\rr_\rmb$ in the presence of an arbitrary 
arrangement of magnetoelectric media. The molecule--field interaction Hamiltonians reduce to
\begin{align}
\label{eq19}
\hat{H}_{\rma F} &= -\hat{\vect{d}}_{\rma}\cdot\hat{\vect{E}}(\vect{r}_{\rma}),\\
\label{eq20}\hat{H}_{\rmb F} &=-\hat{\vect{d}}_{\rmb}\cdot\hat{\vect{E}}(\vect{r}_{\rmb})
-\hat{\vect{m}}_{\rmb}\cdot\hat{\vect{B}}(\vect{r}_{\rmb}).
\end{align}
We first calculate the contribution from every single diagram in Fig.~(\ref{fig1}) in 
the right hand side of Eq.~(\ref{p4}). For example, for diagram (1) in 
Fig.~\ref{fig1}, the respective intermediate states read as follows
\begin{align}
\label{eq21}
|I\rangle &=|m_\rma\rangle|0_\rmb\rangle
|1_{\lambda_1 i_1}(\vect{r}_1,\omega_1)\rangle,\nonumber\\
|I\!I\rangle & = |0_\rma\rangle|0_\rmb\rangle
|1_{\lambda_2i_2}(\vect{r}_2,\omega_2),1_{\lambda_3 i_3}(\vect{r}_3,\omega_3)\rangle, \nonumber\\
|I\!I\!I\rangle & = |0_\rma\rangle|l_\rmb\rangle|1_{\lambda_4 i_4}(\vect{r}_4,\omega_4)\rangle,
\end{align} 
where $|m_\rma\rangle$ and $|l_\rmb\rangle$ denote excited molecular states. The denominator is equal to $\hbar^3(\omega_\rma^{m}+\omega_1)
(\omega_2+\omega_3)(\omega_\rmb^{l}+\omega_4)$ 
with $\omega_{A'}^{i}\equiv(E_{A'}^i-E_{A'}^0)/\hbar$ denoting the transition frequency
of molecule $A'$.
 Substitution of Eq.~(\ref{eq19}) together with Eq.~(\ref{E}) for the electric field operator, making use of the definitions 
(\ref{eq6}) and (\ref{eq7}) for single-- and two--photon states of the electromagnetic field, and applying the commutation relations (\ref{comm1}) and (\ref{comm2}), we find      
\begin{align}
\label{eq22}
\langle I|\hat{H}_{\rma F}|0\rangle=&- \left[\vect{d}_{\rma}^{m0}\cdot\tens{G}^\ast_{\lambda_1}(\vect{r}_\rma,\vect{r}_1,\omega_1)\right]_{i_1},\\
\label{eq23}
\langle I\!I|\hat{H}_{\rma F}|I\rangle =&-\frac{1}{\sqrt{2}}\left[\vect{d}_\rma^{0m}\cdot\tens{G}_{\lambda_2}^\ast(\vect{r}_\rma,\vect{r}_2,\omega_2)\right]_{i_2}\delta^{(31)}\nonumber\\ 
&-\frac{1}{\sqrt{2}}\left[\vect{d}_\rma^{0m}\cdot\tens{G}_{\lambda_3}^\ast(\vect{r}_\rma,\vect{r}_3,\omega_3)\right]_{i_3}\delta^{(21)},
\end{align}
where $\vect{d}_{\rma}^{ij}=\langle i_\rma|\hat{\vect{d}}_{\rma}|j_\rma\rangle$
is the transition electric dipole moment,
 and we have introduced the short--hand notation $\delta^{(\alpha\beta)}$ $\!=$ $\!\delta(\rr_\alpha-\rr_\beta)$ $\!\delta(\omega_\alpha-\omega_\beta)$ $\!
\delta_{i_\alpha i_\beta}$ $\!\delta_{\lambda_\alpha \lambda_\beta}$.
The two photons present in state $|I\!I\rangle$ are to be absorbed by the chiral molecule $\rmb$. Diagram (1) 
refers to two cases of (1)$a$ and (1)$b$ as shown in Fig.~\ref{fig2}, differing in the time--ordering of the electric--dipole and 
magnetic--dipole transitions in the chiral atom $\rmb$.
Restricting our attention to case (1)$a$, similar to Eqs.~(\ref{eq22}) and (\ref{eq23}) we obtain 
\begin{align}
\label{eq24}
\langle I\!I\!I|\hat{H}_{BF}|I\!I\rangle =&-\frac{1}{\sqrt{2}}[\vect{d}_\rmb^{l0}\cdot\tens{G}_{\lambda_2}(\vect{r}_\rmb,\vect{r}_2,\omega_2)]_{i_2}\delta^{(34)} 
-\frac{1}{\sqrt{2}}[\vect{d}_\rmb^{l0}\cdot\tens{G}_{\lambda_3}(\vect{r}_\rmb,\vect{r}_3,\omega_3)]_{i_3}\delta^{(24)},
\\
\label{eq25}
\langle 0|\hat{H}_{\rmb F}|I\!I\!I\rangle=&- \frac{1}{\mi \omega_4}[\vect{m}_{\rmb}^{0l}\cdot\delb\times\tens{G}_{\lambda_4}(\vect{r}_\rmb,\vect{r}_4,\omega_4)]_{i_4}.
\end{align}   
Note that $\hat{H}_{\rmb F}$ is substituted by $-\hat{\vect{d}}_{\rmb}\cdot\hat{\vect{E}}
(\vect{r}_\rmb)$ and
 $-\hat{\vect{m}}_{\rmb}\cdot\hat{\vect{B}}(\vect{r}_\rmb)$, respectively, for the first and second photon absorptions in molecule $\rmb$.

Next we substitute these matrix elements into Eq.~(\ref{p4}).
The summation present in Eq.~(\ref{p4}) runs over discrete variables of molecular states 
$m$, $l$, electric or magnetic nature of the noise source
$\lambda$, spatial components $i$, as well as continuous variables of frequency and 
position.
Making use of the integral relation (\ref{intrel})
leads to the contribution coming from case (1)$a$ to the fourth-order energy shift as
\begin{equation}
\label{eq26}
U^{1a}_{EC}= -\frac{\mu_0^2}{\hbar\pi^2}\sum_{m,l\neq 0} 
\int_0^\infty\dif\omega_1\int_0^\infty\dif\omega_2 \omega_1^2\omega_2^2\frac{N_{EC}(\omega_1,\omega_2)}{D_{1}(\omega_1,\omega_2)},
\end{equation}
where
\begin{align}
\label{eq27}
N_{EC}(\omega_1,\omega_2)&= -\frac{\mi}{\omega_2}\trace
\left[\vect{d}_\rma^{m0}\vect{d}_{\rma}^{0m}\cdot\im
\tens{G}(\vect{r}_\rma,\vect{r}_\rmb,\omega_1)\cdot \vect{d}_{\rmb}^{l0}\vect{m}_\rmb^{0l}\cdot\delb\times\im\tens{G}(\vect{r}_\rmb,\vect{r}_\rma,\omega_2)\right],\\
D_{1}(\omega_1,\omega_2) &= (\omega_\rma^{m}+\omega_1)
(\omega_1+\omega_2)(\omega_\rmb^{l}+\omega_2).
\end{align}
 In writing Eq.~(\ref{eq27}) we have used Lloyd's theorem which allows us to assume that the matrix elements of 
 the electric dipole operator are real-valued while those of the magnetic dipole operator are taken to be pure imaginary; $\vect{d}_{A'}^{ij}=\vect{d}_{A'}^{ji}$, $\vect{m}_{A'}^{ij}=-\vect{m}_{A'}^{ji}$ \cite{0867}.  

The contribution from diagram (1)$b$ can be obtained in a similar manner and the only difference compared with that of 
(1)$a$ is found to be that $N(\omega_1,\omega_2)$ is replaced by $N(\omega_2,\omega_1)$. That is
\begin{align}
\label{eq28}
U^1_{EC}&= U^{1a}_{EC}+U^{1b}_{EC}=-\frac{\mu_0^2}{\hbar\pi^2}\sum_{m,l\neq 0} 
\int_0^\infty\dif\omega_1\int_0^\infty\dif\omega_2 \,\omega_1^2\omega_2^2\frac{N_{EC}(\omega_1,\omega_2)+N_{EC}(\omega_2,\omega_1)}{D_{1}(\omega_1,\omega_2)}\nonumber\\
&=-\frac{\mu_0^2}{\hbar\pi^2}\sum_{m,l\neq 0} 
\int_0^\infty\dif\omega_1\int_0^\infty\dif\omega_2 \omega_1^2\omega_2^2 N_{EC}(\omega_1,\omega_2)\left[\frac{1}{D_1(\omega_1,\omega_2)}+\frac{1}{D_1(\omega_2,\omega_1)}\right].
\end{align} 
For all other diagrams in Fig.~\ref{fig1} the result is seen to be obtained similarly to diagram (1). Summing over all contributions leads to (see \ref{AppA})
\begin{align}
\label{eq33}
U_{EC}&(\rr_\rma,\rr_\rmb)=  \frac{\mu_0^2\mi}{\hbar\pi^2}\sum_{m,l\neq 0}\frac{4\omega_\rma^m}{\omega_\rma^m+\omega_\rmb^l} \int_0^\infty\dif\omega_1
\int_0^\infty\dif\omega_2
\,{\omega_1^2\omega_2}\nonumber\\
&\times\trace\left[\vect{d}_\rma^{m0}\vect{d}_{\rma}^{0m}\cdot\im
\tens{G}(\vect{r}_\rma,\vect{r}_\rmb,\omega_1)\cdot \vect{d}_{\rmb}^{l0}\vect{m}_\rmb^{0l}\cdot\delb\times\im\tens{G}(\vect{r}_\rmb,\vect{r}_\rma,\omega_2)\right]\nonumber\\
&\times\bigg[\frac{1}{(\omega_\rma^m+\omega_2)(\omega_\rmb^l+\omega_2)}-\frac{1}{(\omega_\rma^m+\omega_1)(\omega_\rmb^l+\omega_1)}\bigg]
\bigg(\frac{1}{\omega_2+\omega_1}-\frac{1}{\omega_2-\omega_1}\bigg).
\end{align}

Equation (\ref{eq33}) can be simplified, first, by performing one of the frequency integrals and then rewriting the remaining integral in terms of 
imaginary frequency $\omega=\mi\xi$ (see \ref{AppB}) to yield 
\begin{align}
\label{eq34}
U_{EC}(\vect{r}_\rma,\vect{r}_\rmb)=&-\frac{4\mu_0^2\mi}{\hbar\pi}\int_0^\infty\dif \xi\,\xi^3\sum_{m,l\neq 0}
\frac{\xi\omega_\rma^m}{[(\omega_\rma^m)^2+\xi^2][(\omega_\rmb^l)^2+\xi^2]}\nonumber\\
&\times\trace\left[\vect{d}_\rma^{m0}\vect{d}_{\rma}^{0m}\cdot
\tens{G}(\vect{r}_\rma,\vect{r}_\rmb,\mi \xi)\cdot \vect{d}_{\rmb}^{l0}\vect{m}_\rmb^{0l}\cdot\delb\times\tens{G}(\vect{r}_\rmb,\vect{r}_\rma,\mi \xi)\right].
\end{align} 
This formula can be written as
\begin{align}
\label{eq35}
U_{EC}(\vect{r}_\rma,\vect{r}_\rmb)= &\frac{\hbar\mu_0^2}{\pi}\int_0^\infty\dif \xi\,\xi^3\nonumber\\
&\times \trace\left[
\bm{\alpha}_\rma(\mi \xi)\cdot
\tens{G}(\vect{r}_\rma,\vect{r}_\rmb,\,\mi \xi)\cdot {\bm\chi}^{em}_\rmb
(\mi\xi)\cdot\delb\times\tens{G}(\vect{r}_\rmb,\vect{r}_\rma,\mi \xi)
\right],
\end{align}
 where $\bm{\alpha}(\omega)$ and $\bm{\chi}^{em}(\omega)$ 
 are, respectively, electric polarisability and chiral polariz-
 ability of the molecules, given as
\begin{equation}
\label{alpha}
\bm\alpha_\rma(\omega) =\lim_{\epsilon\to 0^+} \frac{2}{\hbar}\sum_{m} \frac{\omega_\rma^m\vect{d}_\rma^{m0}\vect{d}_\rma^{0m}}{(\omega_\rma^m)^2-\omega^2-\mi \epsilon \omega}\,, 
\end{equation} 
 \begin{equation}
 \label{chiem}
 \bm\chi^{em}_\rmb(\omega) = \lim_{\epsilon\to 0^+}-\frac{2}{\hbar}\sum_{l}
 \frac{\omega\vect{d}_\rmb^{l0}\vect{m}_\rmb^{0l}}{(\omega_\rmb^l)^2-\omega^2-\mi \epsilon\omega}\,.
 \end{equation}
 Equation (\ref{eq35}) agrees with the corresponding result given in 
 Ref.~\cite{Barcellona2017} where the calculation was based upon second-order perturbation theory by introducing an effective two-photon interaction Hamiltonian. 

\begin{figure}[!ht]
	\centering
	\includegraphics[width=.4\linewidth]{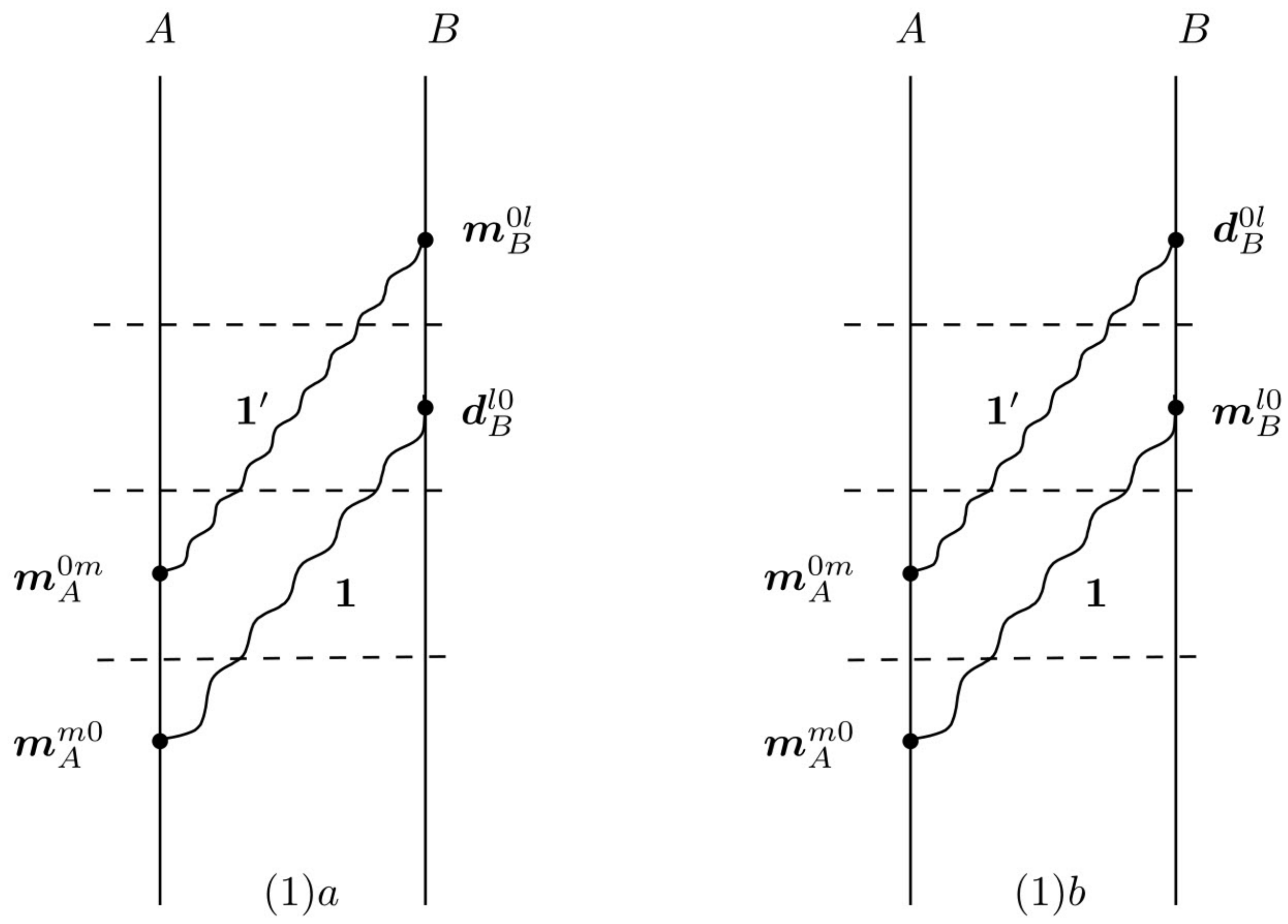}
	\caption{\label{fig3} Diagram $(\mathrm{1})$ in Fig.~\ref{fig1} 
		for the interaction between a\\ paramagnetic molecule $\rma$ and a chiral molecule $\rmb$ being split into two\\ cases, depending on the sequence of the 
		two transitions in the chiral\\ molecule $\rmb$.} 
\end{figure}

\begin{figure}[!ht]
\centering
		\includegraphics[width=.7\linewidth]{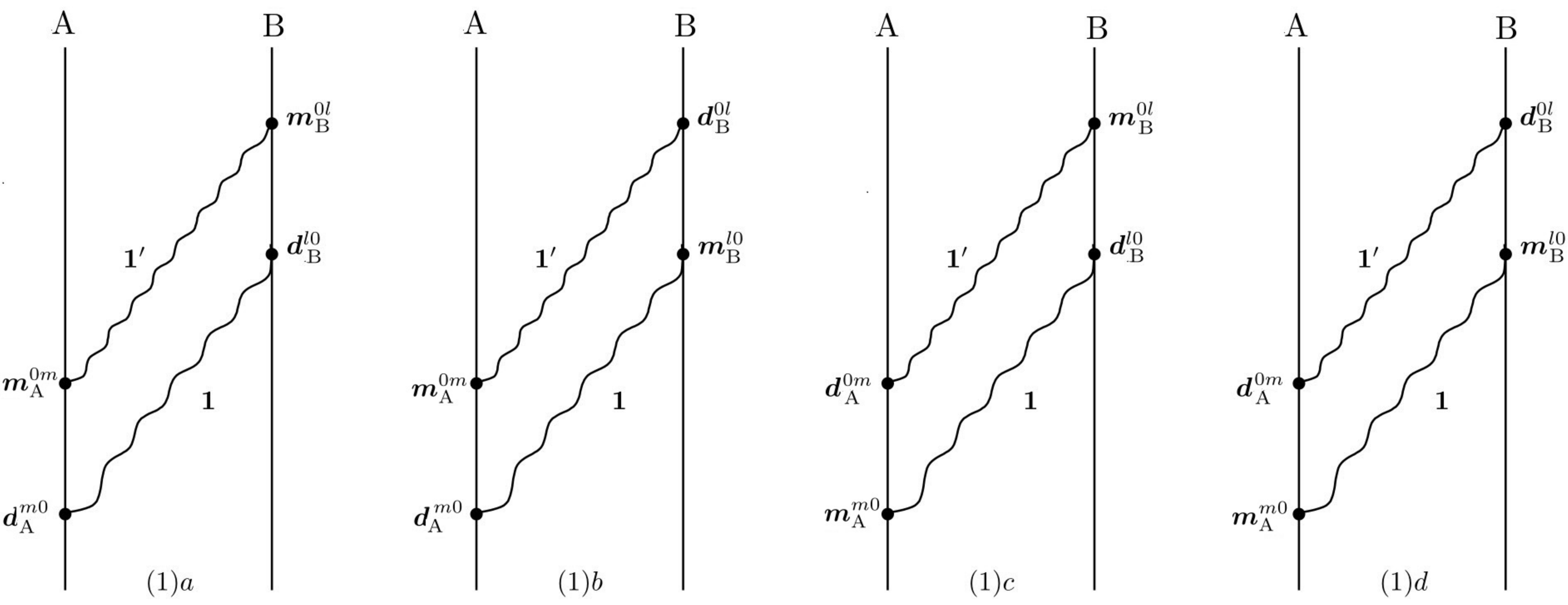}
	\caption{\label{fig4}
	Diagram $(\mathrm{1})$ in Fig.~\ref{fig1} 
		for the interaction between two chiral molecules being split into four cases, depending on the sequence of the
		electric and magnetic transitions in each molecule} 
\end{figure}

\paragraph{ Paramagnetic--chiral interaction.}
By replacing the electric molecule $\rma$ by a paramagnetic one, the interaction Hamiltonian of molecule $\rma$ becomes 
\begin{equation}
\label{Eq38}
\hat{H}_{\rma F} = -\hat{\vect{m}}_{\rma}\cdot\hat{\vect{B}}(\vect{r}_{\rma}),
\end{equation} 
while that of molecule $\rmb$ is again given by Eq.~(\ref{eq20}). Following the same procedure as in obtaining the electric--chiral potential, first, we restrict our attention to case $(1)$ in Fig.~\ref{fig1}, which in turn splits into two cases
$(1)a$ and $(1)b$ shown in Fig.~\ref{fig3} for the paramagnetic--chiral interaction. 
For case $(1)a$ in Fig.~\ref{fig3}, it can be seen easily that the intermediate states and the denominator remain unchanged, the matrix elements $\langle I\!I\!I|\hat{H}_{\rma F}|I\!I\rangle$ and  $\langle 0|\hat{H}_{\rma F}|I\!I\!I\rangle$ are the same as in Eqs.~(\ref{eq24}) and (\ref{eq25}), and the only difference is in the following matrix elements: 
\begin{align}
\label{eq39}
\langle I|\hat{H}_{\rma F}|0\rangle=& \left[\frac{1}{\mi \omega_1}\vect{m}_{\rma}^{m0}\cdot\dela\times
\tens{G}^\ast_{\lambda_1}(\vect{r}_\rma,\vect{r}_1,\omega_1)\right]_{i_1},\\
\label{eq40}
\langle I\!I|\hat{H}_{\rma F}|I\rangle =&\frac{1}{\sqrt{2}}\left[\frac{1}{\mi\omega_2}\vect{m}_\rma^{0m}\cdot\dela\times\tens{G}_{\lambda_2}^\ast(\vect{r}_\rma,\vect{r}_2,\omega_2)\right]_{i_2}\delta^{(31)} \nonumber
\\
&+\frac{1}{\sqrt{2}}\left[\frac{1}{\mi\omega_3}\vect{m}_\rma^{0m}\cdot\dela\times\tens{G}_{\lambda_3}^\ast(\vect{r}_\rma,\vect{r}_3,\omega_3)\right]_{i_3}\delta^{(21)}.
\end{align}
Substitution of these into Eq.~(\ref{p4}) and making use of the integral relation 
(\ref{intrel}) leads for the contribution of diagram $(1)a$ in Fig.~\ref{fig3}, again, to Eq.~(\ref{eq26}) with
$N_{EC}(\omega_1,\omega_2)$ being replaced by
\begin{align}
\label{eq41}
N_{PC}&(\omega_1,\omega_2)=-\frac{\mi}{\omega_1\omega_2^2}\nonumber\\
&
\times\trace\left[\vect{m}_\rma^{m0}\vect{m}_{\rma}^{0m}\!\cdot\!\dela\times\im
\tens{G}(\vect{r}_\rma,\vect{r}_\rmb,\omega_2)\!\times\!\ledb\!\cdot\! 
\vect{m}_{\rmb}^{l0}\vect{d}_\rmb^{0l}\!\cdot\!\im\tens{G}(\vect{r}_\rmb,\vect{r}_\rma,\omega_1)\!\times\!\leda\right].
\end{align}
The contribution of diagram $(1)b$ in Fig.~\ref{fig3} also results in Eq.~(\ref{eq41}) with 
$N_{PC}(\omega_1,\omega_2)$ being replaced by $N_{PC}(\omega_2,\omega_1)$. Taking into account the contribution from all other diagrams in Fig.~\ref{fig1} we obtain (see \ref{AppC})
\begin{align}
\label{eq44}
U_{PC}& (\vect{r}_\rma,\vect{r}_\rmb)= \frac{\mu_0^2\mi}{\hbar\pi^2}\sum_{m,l\neq 0} \frac{4\omega_\rma^m}{\omega_\rma^m+\omega_\rmb^l}\int_0^\infty\dif\omega_1
\int_0^\infty\dif\omega_2
\,{\omega_1}\nonumber\\
&\times\trace\left[\vect{m}_\rma^{m0}\vect{m}_{\rma}^{0m}\cdot\dela\times
\im\tens{G}(\vect{r}_\rma,\vect{r}_\rmb,\omega_2)\times\ledb\cdot 
\vect{m}_{\rmb}^{l0}\vect{d}_\rmb^{0l}\cdot\im\tens{G}(\vect{r}_\rmb,\vect{r}_\rma,\omega_1)\times\leda\right]\nonumber\\
&\times\bigg[\frac{1}{(\omega_\rma^m+\omega_2)(\omega_{\rmb}^l+\omega_2)}-\frac{1}{(\omega_\rma^m+\omega_1)(\omega_\rmb^l+\omega_1)}\bigg]
\bigg(\frac{1}{\omega_2+\omega_1}+\frac{1}{\omega_2-\omega_1}\bigg).
\end{align}
At this stage, in order to present the result in a more convenient form, we may first use the definition (\ref{a3}) and the identity (\ref{a5}) to perform one of the frequency integrals, and then use complex analysis to write the remaining integral in terms of imaginary frequency as outlined in \ref{AppA}. Doing so, we end up with
\begin{align}
\label{eq45}
&U_{PC} (\vect{r}_\rma,\vect{r}_\rmb) =\frac{\hbar\mu_0^2}{\pi}\int_0^\infty\dif \xi\,\xi\nonumber\\
&\times\trace\left[\bm{\beta}_\rma^P(\mi \xi)\cdot\dela\times
\tens{G}(\vect{r}_\rma,\vect{r}_\rmb,\mi \xi)\times\ledb\cdot 
{\bm\chi}_\rmb^{me}(\mi \xi)\cdot\tens{G}(\vect{r}_\rmb,\vect{r}_\rma,\mi \xi)\times\leda\right],
\end{align}
where the paramagnetisability tensor $\bm{\beta}^p_\rma$ and the chiral polarisability ${\bm\chi}_\rmb^{me}$ are defined as
 \begin{align}
 \label{eq46}
 \bm\beta_\rma^P(\omega) &= \lim_{\epsilon\to 0^+}\frac{2}{\hbar}\sum_{m}
 \frac{\omega_\rma^{m}\vect{m}_\rma^{m0}\vect{m}_\rma^{0m}}{(\omega_\rma^{m})^2-\omega^2-\mi \epsilon\omega}\,,\\
 \label{chime}
 \bm\chi^{me}_\rmb(\omega) &= \lim_{\epsilon\to 0^+}-\frac{2}{\hbar}\sum_{l}
 \frac{\omega\vect{m}_\rmb^{l0}\vect{d}_\rmb^{0l}}{(\omega_\rmb^l)^2-\omega^2-\mi \epsilon\omega}\,.
 \end{align}
\paragraph{Chiral--chiral interaction.}
In order to calculate the chiral--chiral contribution to the intermolecular potential, the
molecule--field interaction Hamiltonians that have to be considered are
\begin{align}
\label{eq47}
\hat{H}_{A'F} &=-\hat{\vect{d}}_{A'}\cdot\hat{\vect{E}}(\vect{r}_{A'})
-\hat{\vect{m}}_{A'}\cdot\hat{\vect{B}}(\vect{r}_{A'}), \qquad A'\in\{\rma,\rmb\}
\end{align}
as in each molecule, one of the two opposite transitions (mentioned before) have to be considered of an electric nature with the other being considered a magnetic transition. Hence, every single diagram in Fig.~\ref{fig1} splits into four additional time-orderings. As an example, Fig.~\ref{fig4} shows the resulting split of diagram (1).

Let's consider first, diagram $(1)a$ from Fig.~\ref{fig4}. A calculation similar to the one outlined above Eq.~(\ref{eq22}), leads to the matrix elements $\langle I|\hat{H}_{\rma F}|0\rangle$, 
$\langle I\!I|\hat{H}_{\rma F}|I\rangle$, $\langle I\!I\!I|\hat{H}_{\rma F}|I\!I\rangle$, and $\langle 0|\hat{H}_{\rma F}|I\!I\!I\rangle$ given, respectively, by Eqs.~(\ref{eq22}), (\ref{eq40}), (\ref{eq24}), and (\ref{eq25}).
Substitution of these matrix elements in Eq.~(\ref{p4}), leads to
\begin{equation}
\label{eq48}
U^{(1)a}_{CC}= -\frac{\mu_0^2}{\hbar\pi^2}\sum_{m,l\neq 0} 
\int_0^\infty\dif\omega_1\int_0^\infty\dif\omega_2 \omega_1^2\omega_2^2\frac{N^{1a}_{CC}(\omega_1,\omega_2)}{D_{1}(\omega_1,\omega_2)}\,,
\end{equation}
where
\begin{align}
\label{eq49}
N^{1a}_{CC}&(\omega_1,\omega_2) \nonumber\\
&= \frac{1}{\omega_2^2}\trace\left[\vect{d}_\rma^{m0}\vect{m}_{\rma}^{0m}\cdot\dela\times\im
\tens{G}(\vect{r}_\rma,\vect{r}_\rmb,\omega_2)\times\overleftarrow{\nabla}_\rmb\cdot 
\vect{m}_{\rmb}^{l0}\vect{d}_{\rmb}^{0l}\cdot\im\tens{G}(\vect{r}_\rmb,\vect{r}_\rma,\omega_1)\right].
\end{align}
The contributions $U^{(1)b}$, $U^{(1)c}$, and 
$U^{(1)d}$ are found to have the same form as the right hand side of Eq.~(\ref{eq48}),
with $N^{1a}_{CC}$ being replaced, respectively, by $N^{1b}_{CC}$, $N^{1c}_{CC}$, and  $N^{1d}_{CC}$,
where
\begin{align}
\label{eq50}
N^{1b}_{CC}&(\omega_1,\omega_2)\nonumber\\
&=-\frac{1}{\omega_1\omega_2}\trace\left[\vect{d}_\rma^{m0}
\vect{m}_{\rma}^{0m}\cdot\dela\times
\im\tens{G}(\vect{r}_\rma,\vect{r}_\rmb,\omega_2)
\cdot\vect{d}_\rmb^{l0}\vect{m}_{\rmb}^{0l}\cdot\delb\times\im\tens{G}(\vect{r}_\rmb,\vect{r}_\rma,\omega_1)
\right],\nonumber\\
N^{1c}_{CC}&(\omega_1,\omega_2) = N^{1b}_{CC}(\omega_2,\omega_1),\qquad
N^{1d}_{CC}(\omega_1,\omega_2) = N^{1a}_{CC}(\omega_2,\omega_1).
\end{align}
Using these, the contribution of diagram (1) in Fig.~\ref{fig1} to the chiral--chiral interaction potential becomes
\begin{align}
\label{eq51}
U^{(1)}_{CC}=&-\frac{\mu_0^2}{\hbar\pi^2}
\sum_{m,l\neq 0}\int_0^\infty\dif\omega_1\int_0^\infty\dif\omega_2 \,
\omega_1^2\omega_2^2 \nonumber\\
&\times\left[N_{CC}^{1a}(\omega_1,\omega_2)+N_{CC}^{1b}(\omega_1,\omega_2)\right]
\left[\frac{1}{D_1(\omega_1,\omega_2)}+\frac{1}{D_1(\omega_2,\omega_1)}\right].
\end{align}
By calculating the contribution of all other diagrams in Fig.~\ref{fig1} in a similar manner 
and summing them up, we obtain (see \ref{AppD}) the 
chiral--chiral interaction potential as 
\begin{eqnarray}
\label{eq58}
&&\hspace{-1in}U_{CC}(\vect{r}_\rma,\vect{r}_\rmb)=-\frac{\hbar\mu_0^2}{\pi}\int_0^\infty \dif \xi\,\xi^2\nonumber\\
&&\hspace{-.5in}\times\Big\{\trace\left[\bm{\chi}^{em}_\rma(\mi \xi)\cdot\dela\times
\tens{G}(\vect{r}_\rma,\vect{r}_\rmb,\mi \xi)\times\overleftarrow{\nabla}_\rmb
\cdot\bm{\chi}^{me}_\rmb(\mi \xi)\cdot\tens{G}(\vect{r}_\rmb,\vect{r}_\rma,\mi \xi)\right]\nonumber\\
&&\hspace{-.2in}+\trace\left[\bm{\chi}^{em}_\rma(\mi \xi)\cdot\dela\times
\tens{G}(\vect{r}_\rma,\vect{r}_{\rmb},\mi \xi)
\cdot\bm{\chi}^{em}_\rmb(\mi \xi)\cdot\delb\times\tens{G}(\vect{r}_\rmb,\vect{r}_\rma,\mi \xi)
\right]\Big\}.
\end{eqnarray}

\subsection{Diamagnetic--chiral interaction}
\begin{figure}[!ht]
\centering
		\includegraphics[width=.6\linewidth]{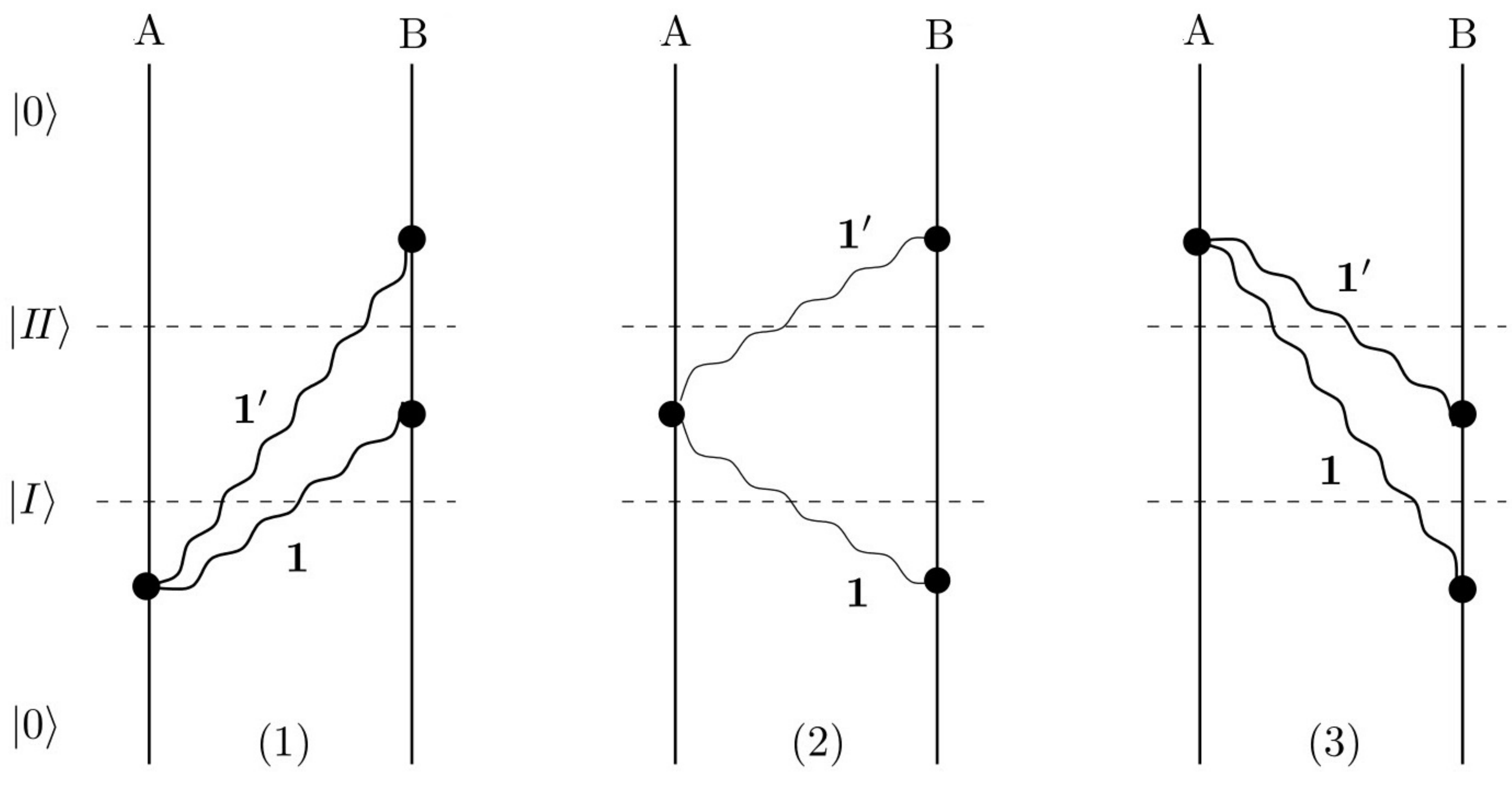}
	\caption{\label{fig5}
	Diagrams for obtaining the interaction potential between a diamagnetic molecule $\rma$ and a chiral molecule $\rmb$ via Eq.~(\ref{p3}).}
\end{figure}

While obtaining the formula of interaction potentials so far required 
fourth-order perturbation theory, the 
diamagnetic--chiral interaction potential is obtained via third-order perturbation theory (cf.~Ref.~\cite{buhmann2013-3}), 
\begin{equation}
\label{p3}
U_{DC}(\rr_\rma,\rr_\rmb)=
\sum_{I,I\!I\neq 0}
\hspace{-1ex}
\frac{\langle 0 |\hat{H}_{\mathrm{int}}|I\!I\rangle\langle I\!I|\hat{H}_{\mathrm{int}}|I\rangle
	\langle I|\hat{H}_{\mathrm{int}}|0\rangle}
{(E_{I\!I}-E_0)(E_{I}-E_0)}\,.
\end{equation}
The underlying reason is that the diamagnetic molecule--field interaction is a two--photon process. The two photons are 
exchanged with the chiral molecule to participate in the two--molecule interaction. 
To perform the calculation, we replace the paramagnetic molecule $\rma$ 
in the preceeding section with a diamagnetic one. 
That is
\begin{equation}
\hat{H}_{\rma F} = -\frac{1}{2}\hat{\vect{B}}(\rr_\rma)
\cdot
\hat{\bm{\beta}}_\rma^d\cdot \hat{\vect{B}}(\rr_\rma),
\end{equation}
 with $\hat{\bm{\beta}}_\rma^d$ defined in Eq.~(\ref{beta}), and $\hat{H}_{\rmb F} $ was given by Eq.~(\ref{eq20}). 
 The complete set of intermediate states that are involved in the perturbation formula (\ref{p3}) 
can be taken into account by the diagrams given in Fig.~\ref{fig5} (see Ref.~\cite{buhmann2013-3}). 

\begin{figure}[!ht]
\centering
		\includegraphics[width=.4\linewidth]{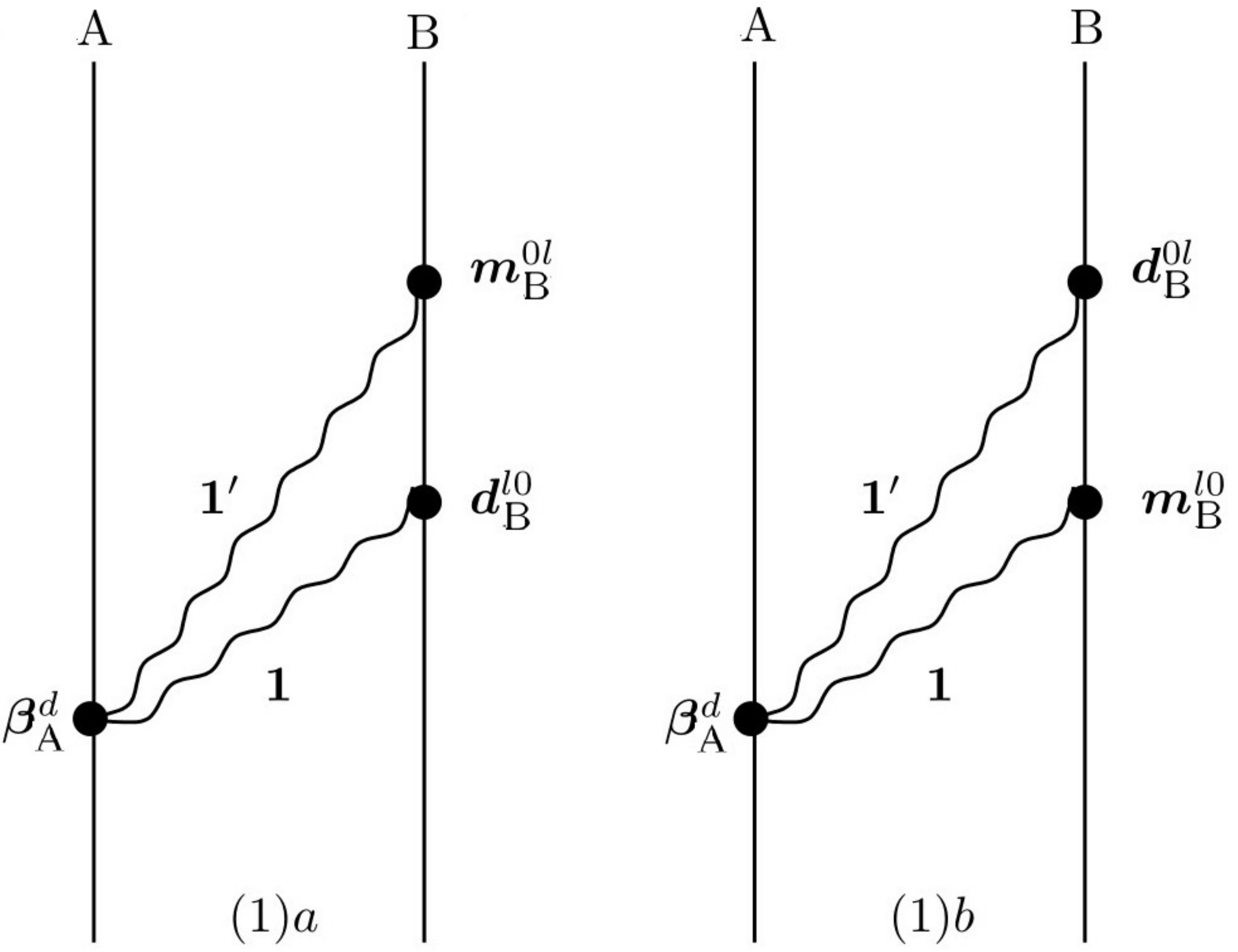}
	\caption{\label{fig6}
	Diagram (1) from Fig.~\ref{fig5} being split into two cases depending on the sequence of the transitions in the chiral molecule $\rmb$.} 
\end{figure}

Each one of the three diagrams in Fig.~\ref{fig5} is split, as shown in Fig.~\ref{fig6}, into two cases depending on whether the first
 transition occurring in the chiral molecule $\rmb$ is of electric type or magnetic type, labeled by $a$ or $b$, respectively, as shown in Fig.~\ref{fig6}.

Let's consider first diagram (1) in Fig.~\ref{fig5}. The corresponding intermediate states are given as follows
\begin{align}
|I\rangle &= |0_\rma\rangle|0_\rmb\rangle|1_{\lambda_1 i_1}(\rr_1,\omega_1),1_{\lambda_2 i_2}(\rr_2,\omega_2)\rangle,\nonumber\\
|I\!I\rangle &= |0_\rma\rangle|l_\rmb\rangle|1_{\lambda_3 i_3}(\rr_3,\omega_3)\rangle.
\end{align}
To detemine the corresponding matrix elements present in the numerator of Eq.~(\ref{p3}), 
diagram (1)$a$ or (1)$b$ in Fig.~\ref{fig6} have to be distinguished. For the former one finds
\begin{align}
\langle I|\hat{H}_{\mathrm{int}}|0\rangle&= -\frac{1}{2}\beta^D_{\alpha\beta}\langle 1_{\lambda_1 i_1}(\rr_1,\omega_1),1_{\lambda_2 i_2}(\rr_2,\omega_2)|\hat{B}_\alpha(\vect r_\rma)\hat{B}_\beta(\vect r_\rma)
|\{0\}\rangle
\nonumber\\
&
= \frac{\beta^D_{\alpha\beta}}{\sqrt{2}\omega_1\omega_2}
\big[\dela\times\tens{G}_{\lambda_1}
(\vect{r}_\rma,\vect{r}_1,\omega_1)\big]_{\alpha i_1}\big[\dela\times\tens{G}_{\lambda_2}
(\vect{r}_\rma,\vect{r}_2,\omega_2)\big]_{\beta i_2},
\end{align}
\begin{align}
\langle I\!I|\hat{H}_{\mathrm{int}}|I\rangle&=\langle 0_\rma|\langle l_\rmb|
\langle 1_{\lambda_3 i_3}(\rr_3,\omega_3)|
-\hat{\vect{d}}_{\rmb}\cdot\hat{\vect{E}}(\vect{r}_{\rmb})
|0_\rma\rangle|0_\rmb\rangle|1_{\lambda_1 i_1}(\rr_1,\omega_1),1_{\lambda_2 i_2}(\rr_2,\omega_2)\rangle\nonumber\\
&= 
-\frac{d_{\rmb}^{l0}}{\sqrt{2}}
\big[\tens{G}_{\lambda_1}
(\vect{r}_\rmb,\vect{r}_1,\omega_1)\big]_{\gamma i_1}\delta^{(32)}
-\frac{d_{\rmb\gamma}^{l0}}{\sqrt{2}}
\big[\tens{G}_{\lambda_2}
(\vect{r}_\rmb,\vect{r}_2,\omega_2)\big]_{\gamma i_2}\delta^{(31)} ,
\end{align}
\begin{align}
\langle 0|\hat{H}_{\mathrm{int}}|I\!I\rangle
&=\langle 0|-\frac{1}{2}\hat{\vect{m}}_\rmb\cdot\hat{\vect{B}}(\vect{r}_\rmb)|I\!I\rangle
=-\frac{m_\eta^{0l}}{\mi \omega_3}
\big[\delb\times\tens{G}_{\lambda_3}
(\vect{r}_\rmb,\vect{r}_3,\omega_3)\big]_{\eta i_3},
\end{align}
with the respective denominator in Eq.~(\ref{p3}) being $\hbar^2(\omega_3$ $\!+\omega_\rmb^{l})(\omega_1$ $\!+\omega_2)$. After substituting these matrix elements in Eq.~(\ref{p3}), the implicitly included summations over $\lambda_j$ ($j=e,m$) and position integrals can be performed on making use of 
the integral relation (\ref{intrel}) to lead to the contribution from diagram (1)$a$ in
Fig.~\ref{fig6} as
\begin{align}
\label{eq65}
&\hspace{-2ex}U^{1a}_{DC} =\frac{\mu_0^2\mi}{\pi^2}\sum_l
\int_0^\infty\dif\omega_1\int_0^\infty\dif\omega_2 {D}_{1a}(\omega_1,\omega_2)
\nonumber\\
&\times
\trace\big\{
\bm{\beta}^D\cdot\dela\times\im\tens{G}(\vect{r}_\rma,\vect{r}_\rmb,\omega_2)\times\ledb\cdot\vect{m}_\rmb^{l0}\vect{d}_\rmb^{0l}
\cdot\big[\im\tens{G}(\vect{r}_\rmb,\vect{r}_\rma,\omega_1)\times\leda\big]
\big\},
\end{align}
where
\begin{equation}
{D}_{1a}(\omega_1,\omega_2)=
\frac{\omega_1 }{(\omega_1+\omega_2)(\omega_\rmb^l+\omega_2)}.
\end{equation}
The contributions from each diagram, $U^{ia(b)}_{DC}$,
can be obtained similarly. It is found that the only difference is in ${D}^{ia(b)}(\omega_1,\omega_2)$
 as listed below
 \begin{equation}
 \label{eq67}
 {D}_{1b}={D}_{3a}=\frac{-\omega_1}{(\omega_1+\omega_2)(\omega_\rmb^l+\omega_1)},
 \,\,
 {D}_{2a}={D}_{2b}= \frac{-\omega_1}{(\omega_\rmb^l+\omega_1)(\omega_\rmb^l+\omega_2)},\,\,{D}_{3b}={D}_{1a}.
 \end{equation}
Using these for $U^{ia(b)}$ and summing up the resulting expressions, as shown in \ref{AppE}, leads to the 
diamagnetic--chiral interaction potential as 
\begin{align}
\label{eq68}
U_{DC}&(\rr_\rma,\rr_\rmb) = \frac{\hbar\mu_0^2}{\pi}\int_0^\infty\dif\xi\,\xi
\nonumber\\
&\times
\trace\big\{
\bm{\beta}^D_\rma\cdot\big[\dela\times\tens{G}(\vect{r}_\rma,\vect{r}_\rmb,\mi \xi)\times\ledb\big]\cdot \bm{\chi}^{me}_\rmb(\mi \xi)
\cdot\big[\tens{G}(\vect{r}_\rmb,\vect{r}_\rma,\mi \xi)\times\leda\big]\big\}.
\end{align}
It is worth noting that the formula for the diamagnetic--chiral interaction potential and that for the paramagnetic--chiral energy shift, Eqs.~(\ref{eq45}) and (\ref{eq68}), despite being obtained from different orders of perturbative calculations, have exactly the same form. These may be summed in a single formula
by substitution of $\bm{\beta}_P(\omega)+\bm{\beta}_D\to \bm{\beta}(\omega)$ to give an overall magnetic--chiral interaction,
\begin{align}
\label{eq69}
U_{MC}&(\rr_\rma,\rr_\rmb) = \frac{\hbar\mu_0^2}{\pi}\int_0^\infty\dif \xi\, \xi
\nonumber\\
&\times
\trace\big\{
\bm{\beta}_\rma(\mi \xi)\cdot\big[\dela\times\tens{G}(\vect{r}_\rma,\vect{r}_\rmb,\mi \xi)\times\ledb\big]\cdot \bm{\chi}^{me}_\rmb(\mi \xi)
\cdot\big[\tens{G}(\vect{r}_\rmb,\vect{r}_\rma,\mi \xi)\times\leda\big]\big\}.
\end{align}

The chiral polarisabilities of the two enantiomers of a chiral molecule oppose each other in algebraic sign. This results in a discriminatory vdW dispersion interaction for the two enantiomers as it can be seen from the expressions obtained, Eqs.~\eqref{eq35}, \eqref{eq45}, \eqref{eq58}, and \eqref{eq68}, where a sign change emerges as a chiral molecule is replaced by its enantiomer.

\section{Duality Invariance}
\label{duality}
The formulas for various contributions to the vdW dispersion interaction given so far in this paper 
with one or both ground-state molecules being chiral, complete the formulas given
for non-chiral species in literature (see Refs.~\cite{safari2006, buhmann2013-3, safari2008}). In this section
we gather all contributions and discuss their symmetry with respect to the duality of electric and magnetic fields. 

As a preparation, we note that all vdW interactions presented in this paper except for the diamagnetic one derive from the bilinear interaction 
\begin{equation}
\label{HA'2}
\hat{H}_{A'F} = -\hat{\vect{d}}_{A'}\cdot\hat{\vect{E}}(\vect{r}_{A'})
-\hat{\vect{m}}_{A'}\cdot\hat{\vect{B}}(\vect{r}_{A'}).
\end{equation} 
Recall that for $\vect{r}_{A'}$ in free space, the electric and magnetic fields featuring in this interaction obey duality invariance \cite{Buhmann2009b,buhmann2009}: when combining a known solution to the Maxwell equation into a dual pair $\vect{E}_\lambda$ ($\lambda=e,m$) with $\vect{E}_e=\vect{E}$, $\vect{E}_m=c\vect{B}$, then applying a rotation in this two-dimensional duality space
\begin{equation}
\begin{pmatrix}\vect{E}_e^\star\\ 
 \vect{E}_m^\star\end{pmatrix} 
=\mathcal{D}(\theta)
\begin{pmatrix}\vect{E}_e\\ 
\vect{E}_m\end{pmatrix}, 
 \quad
 \mathcal{D}(\theta)
 =\begin{pmatrix}\cos\theta&\sin\theta\\ 
 -\sin\theta&\cos\theta\end{pmatrix}
\end{equation}
generates a new solution to the Maxwell equations. It follows immediately that the bilinear interaction Hamiltonian above can be written in duality-invariant form
\begin{equation}
\label{HA'3}
\hat{H}_{A'F} = -\sum_\lambda\hat{\vect{d}}{}_{A'}^\lambda\cdot\hat{\vect{E}}_\lambda(\vect{r}_{A'})
\end{equation} 
by introducing a dual pair vector of molecular dipoles $\vect{d}^\lambda$ with $\vect{d}^e=\vect{d}$, $\vect{d}^m=\vect{m}/c$ and  positing that such an upper-index dual pair vector transforms with $\mathcal{D}^{-1}(\theta)=\mathcal{D}^\trans(\theta)$ under a duality rotation.

If we now calculate the vdW potential arising from this molecule--field interaction using dual-pair notation, then we arrive at a total interaction
\begin{equation}
\label{Uu}
U(\rr_\rma,\rr_\rmb) =  \sum_{\lambda_1,\lambda_2,\lambda_3,\lambda_4= e,m}
U_{\lambda_1\lambda_2\lambda_3\lambda_4}
\end{equation}
with
\begin{multline}
  \label{eq82}
U_{\lambda_1\lambda_2\lambda_3\lambda_4}(\vect{r}_\rma,\vect{r}_\rmb)=  -\frac{\hbar}{2\pi\varepsilon_0^2}\int_0^\infty\dif \xi \\
\times \trace\left[
\bm{\alpha}^{\lambda_1\lambda_2}_\rma(\mi \xi )\cdot
\tens{G}_{\lambda_2\lambda_3}(\vect{r}_\rma,\vect{r}_\rmb,\,\mi \xi )\cdot {\bm\alpha}_\rmb^{\lambda_3\lambda_4}
(\mi \xi )\cdot\tens{G}_{\lambda_4\lambda_1}(\vect{r}_\rmb,\vect{r}_\rma,\mi \xi )
\right]
\end{multline}
which is duality-invariant given that both atoms are situated in a free-space region. Here, we have introduced Green tensor components in duality space
\begin{align}
\label{eq73}
\tens{G}_{ee}(\rr,\rr',\omega) &\equiv \frac{\mi \omega}{c}\tens{G}(\rr,\rr',\omega)\biggl(-\frac{\mi \omega}{c}\biggr),\\
 \label{eq74}
 \tens{G}_{em}(\rr,\rr',\omega)&\equiv \frac{\mi \omega}{c}\tens{G}(\rr,\rr',\omega)\times\bigl(-\overleftarrow\nabla'\bigr),\\
\label{eq75}
 \tens{G}_{me}(\rr,\rr',\omega)&\equiv \nabla\times\tens{G}(\rr,\rr',\omega)\biggl(-\frac{\mi \omega}{c}\biggr),\\
 \label{eq76}
 \tens{G}_{mm}(\rr,\rr',\omega)&\equiv \nabla\times\tens{G}(\rr,\rr',\omega)\times\bigl(-\overleftarrow\nabla'\bigr),
\end{align}
which emerge from expectation values $\langle\hat{\vect{E}}_\lambda\hat{\vect{E}}^\dagger_\lambda\rangle$ and transform via 
$\mathcal{D}(\theta)$ as well as polarisabilities
\begin{align}
\label{eq77}
\bm\alpha^{ee}(\omega)&\equiv\bm{\alpha}(\omega) ,\\
 \label{eq78}
 \bm\alpha^{em}(\omega)&\equiv\bm\chi^{em}(\omega)/c,\\
 \label{eq79}
 \bm\alpha^{me}(\omega)& \equiv\bm\chi^{me}(\omega)/c,\\
 \label{eq80}
 \bm\alpha^{mm}(\omega)&\equiv\bm{\beta}(\omega)/c^2
\end{align}
which transform via $\mathcal{D}^\trans(\theta)$. Equations~(\ref{Uu}) and (\ref{eq82}) give the most general vdW potential of two molecules with dipole polarisabilities, magnetisabilities, and electromagnetic cross-polarisabilities within an arbitrary bi-anisotropic environment. By virtue of $\bm{\beta}(\omega)=\bm{\beta}_P(\omega)+\bm{\beta}_D$, this includes molecules with both paramagnetic and diamagnetic responses. The general potential is duality invariant, provided that both molecules are situated in free-space regions. For molecules in media, duality-invariance can be ensured by using the real-cavity model where the molecules are surrounded by small free-space cavities \cite{Sambale2007}.

The interactions considered in the previous sections emerge by identifying $E=ee$, $M=mm$, $C=em+me$:
\begin{gather}
\label{eq81}
U_{EE}=U_{eeee},\quad U_{EM}=U_{eemm},\quad U_{ME}=U_{mmee},\quad U_{MM}=U_{mmmm},\\
\label{eq84}
U_{EC} =U_{eeem}+U_{eeme},\quad U_{MC} =U_{mmem}+U_{mmme},\\
\label{eq81b}
U_{CE} =U_{emee}+U_{meee},\quad U_{CM} =U_{emmm}+U_{memm},\\
\label{eq84b}
U_{CC} =U_{emem}+U_{emme}+U_{meem}+U_{meme}
\end{gather}
where for chiral responses, one may use Lloyd's theorem
$\bm\chi^{me}=-{{\bm\chi}^{em}}^\top$ [compare Eqs.~(\ref{chiem}) and (\ref{chime})] \cite{0867}. 

The duality invariance with $\theta=\pi/2$ can be exploited for molecules with electric, chiral, and paramagnetic reponses by replacing $\bm{\alpha} \leftrightarrow \bm{\beta}/c^2$ and $\bm{\chi}^{em} \leftrightarrow -\bm{\chi}^{me}$, thus generating new potentials from previously calculated results. For instance, this explains why $U_{EE}$ and $U_{MM}$ in free space are so strikingly similar and the same holds for $U_{EC}$ and $U_{MC}$. For diamagnetic molecules, this transformation applies only formally, since the diamagnetic magnetisability is negative, in contrast to the electric polarisability from which it is obtained via the transformation. 

One may be tempted to generate chiral potentials from electric or magnetic ones by applying a duality transformation $\theta=\pi/2$. However, as noted in Refs.~\cite{buhmann2012,Buhmann2018charge}, such a transformation will generate nonreciprocal cross-polarisabilities which do not obey Lloyd's theorem and hence do not correspond to chiral molecules, but instead charge-parity violating ones.  For instance, starting from a purely electric isotropic molecule of polarisability $\alpha$, the transformed molecule will exhibit cross-polarisabilities $\chi^{em}/c=\chi^{me}/c=\alpha/4$ which clearly violate Lloyd's theorem. This is why vdW potentials involving chiral molecules have a distance dependence which is quite distinct from the known dependences for electric and magnetic molecules. We will see this explicitly below when studying free-space examples.

\section{Interactions in Free Space}
\label{freespace}
As the simplest application of the obtained formulas, we may consider two molecules in free space for which it can easily be verified that the chirality contribution vanishes
in the case of isotropic molecules
 unless both molecules are chiral. Hence, for the sake of generality we consider anisotropic molecules and treat the case of isotropic molecules as specific examples. For notational convenience we replace, in what follows, $\bm{\chi}^{em}$ with $\bm{\chi}$ and use the fact that $\bm{\chi}^{me}=-\bm{\chi}^{em\,\top} = -\bm{\chi}^{\top}$.
 
\subsection{Electric--chiral interaction}
\label{EC-free}
The free space Green's tensor $\tens{G}^{(0)}$ reads \cite{buhmann2013-1}
\begin{equation}
\label{G0}
\tens{G}^{(0)}(\rr_\rma,\rr_\rmb,\mi \xi) = \frac{1}{4\pi}
\left(\tens{I} -\frac{c^2}{\xi^2}\nabla_\rma\nabla_\rma\right)\frac{\me^{- R\xi/c}}{R} = \frac{\me^{-kR}}{4\pi k^2 R^3}\left[ f(kR)\tens{I} - g(kR) \frac{\vect{R}\vect{R}}{R^2} \right],
\end{equation}
where $k= \xi/c$,  $R=|\vect{R}|$, $\vect{R}  = \rr_\rma - \rr_\rmb$, and
\begin{align}
\label{f&g}
f(x)=&1+x+x^2,\nonumber\\
g(x)=&3+3x+x^2 .
\end{align}
Using this tensor together with the required derivative 
\begin{equation}
\label{curlG}
\nabla_\rmb\times \tens{G}^{(0)}(\rr_\rmb,\rr_\rma,\mi \xi)
= \frac{1}{4\pi}\nabla_\rmb\left(\frac{\me^{- R\xi/c}}{R}\right)\times\tens{I}
=\frac{\me^{-R\xi/c}}{4\pi R^3}(1+ R\xi/c)\vect{R}\times\tens{I}
\end{equation}
in Eq.~(\ref{eq35}) for an electric molecule A and a chiral molecule
B, results in
\begin{multline}
\label{UECfree}
U_{EC}(\vect{R})=\frac{\hbar}{16 \pi^3 \varepsilon_0^2} \epsilon_{ipq}\tilde{R}_q \int_0^\infty \dif k \,k^6 \me^{-2kR}  \alpha_A^{ij}(\mi kc)  \chi_B^{rp}
(\mi kc) \\
\times \left[ \frac{\delta_{jr} -\tilde{R}_j \tilde{R}_r}{k^2R^2} + \frac{2 \left(\delta_{jr} -2\tilde{R}_j \tilde{R}_r \right)}{k^3R^3}  +\frac{ 2 \left(\delta_{jr} -3 \tilde{R}_j \tilde{R}_r \right)}{k^4R^4} + \frac{\delta_{jr} -3\tilde{R}_j \tilde{R}_r}{k^5R^5} \right]
\end{multline}
with $\tilde{\vect{R}} = \vect{R}/R$. In the case of isotropic molecules, for which $\bm{\alpha}$ and $\bm{\chi}$ are diagonal matrices, it can be seen easily that the interaction potential vanishes.

In the retarded limit of intermolecular separation ($R\gg c/\omega_{\mathrm{A}}$ with $\omega_{\mathrm{A}}$ being a typical molecular transition frequency), $\bm{\alpha}$ and $\bm{\chi}$ in Eq.~\eqref{UECfree} 
can be replaced by their static values $\bm{\alpha}(0)$ and $\bm{\chi}(0)$,
\begin{equation}
\label{alpha0}
\bm{\alpha}_{\mathrm{A}}(0) = \frac{2}{\hbar}\sum_m\frac{\vect{d}_A^{m0}\vect{d}_A^{0m}}{\omega_A^{m}}\equiv \bm{\alpha}_{\mathrm{A}}\,,
\qquad \lim_{k\to 0}\bm{\chi}_{\mathrm{B}}(\mi k c) = \frac{-2\mi}{\hbar}kc\sum_l\frac{\vect{d}_B^{l0}\vect{m}_B^{0l}}{\omega_B^{l}}
\equiv k\bm{\chi}'_{\mathrm{B}}.
\end{equation}
Performing the remaining integral leads to
\begin{equation}
\label{UECfreeRet}
U_{EC}^{\text{r}}(\vect{R})=\frac{7 \hbar}{128 \pi^3 \varepsilon_0^2 R^8}  \epsilon_{ipq} \tilde{R}_q \left(5\delta_{jr} -9 \tilde{R}_j \tilde{R}_r \right) \alpha_A^{ij}  \chi{'}_B^{rp}.
\end{equation}
In the opposite limit of non-retarded coupling, $R\ll c/\omega_{\rma}$, the exponential factor in Eq.~\eqref{UECfree} tends to unity and the last term in the square brackets gives the main contribution. 
Using the definitions \eqref{alpha} and \eqref{chiem} into Eq.~\eqref{UECfree} and carrying out the integral  we find
\begin{equation}
\label{UECnonret}
U_{EC}^{\mathrm{nr}}(\vect{R})=-\frac{\mi\mu_0}{8 \pi^2 \varepsilon_0\hbar R^5} \epsilon_{ipq} \tilde{R}_q \left(\delta_{jr} -3 \tilde{R}_j \tilde{R}_r \right) \sum_{l,m} \left( \vect{d}_A^{m0}\vect{d}_A^{0m} \right)^{ij}
 \left( \vect{d}_B^{l0}\vect{m}_B^{0l} \right)^{rp} \frac{\omega_A^m}{\omega_A^m+\omega_B^l}\,.
\end{equation}

\subsection{Magnetic--chiral interaction}
\label{MC-free}
The interaction potential of a magnetic molecule A and a chiral molecule
B in free space may be written by direct calculation of the required derivatives of the Green's tensor,
\begin{equation}
\label{curlGcurl}\dela\times \tens{G}^{(0)}(\rr_\rma,\rr_\rmb,\omega)\times\ledb
=-\frac{\omega^2}{c^2}\tens{G}^{(0)}(\rr_\rma,\rr_\rmb,\omega),
\end{equation}
and
\begin{equation}
\label{Gcurl}
 \tens{G}^{(0)}(\rr_\rmb,\rr_\rma,\omega)\times\leda
= -\delb\times \tens{G}^{(0)}(\rr_\rmb,\rr_\rma,\omega),
\end{equation}
together with Eq.~\eqref{curlG}, and using them in Eq.~\eqref{eq69}. Also, it can be written easily on making use of the duality transformation mentioned below Eq.~\eqref{eq84} as
\begin{align}
\label{UMCfree}
 &U_{MC}(\vect{R}) = \frac{\hbar\mu_0^2c^2}{16 \pi^3} \epsilon_{ipq}\tilde{R}_q \int_0^\infty \dif k \,k^6 \me^{-2kR}  \beta_A^{ij}(\mi kc)  \chi_B^{pr}
(\mi kc) \nonumber\\
&\times \left[ \frac{\delta_{jr} -\tilde{R}_j \tilde{R}_r}{k^2R^2} + \frac{2 \left(\delta_{jr} -2\tilde{R}_j \tilde{R}_r \right)}{k^3R^3}  +\frac{ 2 \left(\delta_{jr} -3 \tilde{R}_j \tilde{R}_r \right)}{k^4R^4} + \frac{\delta_{jr} -3\tilde{R}_j \tilde{R}_r}{k^5R^5} \right].
\end{align}
Similar to the electric--chiral interaction potential in free space, the magnetic--chiral one vanishes for isotropic molecules by orientational averaging.

In the retarded limit, the result is obtained through a similar manner as for the electric--chiral interacton potential, as
\begin{equation}
\label{UMCfreeRet}
U_{MC}^{\text{r}}(\vect{R})=\frac{7 \hbar\mu_0^2c^2}{128 \pi^3  R^8}  \epsilon_{ipq} \tilde{R}_q \left(5\delta_{jr} -9 \tilde{R}_j \tilde{R}_r \right) \beta_A^{ij}  \chi{'}_B^{pr}\,,
\end{equation}
where 
\begin{equation}
\bm{\beta}_\rma = \bm{\beta}^d_{\mathrm{A}}+\bm{\beta}^p_{\mathrm{A}}(0) = \bm{\beta}^d_{\mathrm{A}}+\frac{2}{\hbar}\sum_m\frac{\vect{m}_A^{m0}\vect{m}_A^{0m}}{\omega_A^{m}} \,.
\end{equation}

In the non-retarded limit, the diamagnetic--chiral and paramagnetic--chiral interactions have to be treated separately because of the frequency-independent nature of the diamagnetisability. For the paramagnetic--chiral interaction, through a similar discussion for electric--chiral contribution that led to Eq.~\eqref{UECfreeRet}, we obtain
\begin{equation}
\label{UPCnonret}
U_{PC}^\mathrm{nr}(\vect{R})=-\frac{\mi\mu_0^2}{8 \pi^2 \hbar R^5} \epsilon_{ipq} \tilde{R}_q \left(\delta_{jr} -3 \tilde{R}_j \tilde{R}_r \right) \sum_{l,m}  \frac{\omega_A^m \left( \vect{m}_A^{m0}\vect{m}_A^{0m} \right)^{ij}
 \left( \vect{d}_B^{l0}\vect{m}_B^{0l} \right)^{pr}}{\omega_A^m+\omega_B^l}\,.
\end{equation}
For the diamagnetic--chiral interaction, on the contrary, the major contribution to the frequency integral comes from frequencies much greater than $c/\omega_{A}$, for which the chiral polarisability, Eq.~\eqref{chiem}, approximates to
 \begin{equation}
 \label{eq87}
 \bm\chi^{em}_\rmb(\omega) = \frac{2}{\hbar\omega}\sum_{l}
\vect{d}_\rmb^{l0}\vect{m}_\rmb^{0l}\,.
 \end{equation}
Using Eq.~(\ref{eq87}) in Eq.~\eqref{UMCfree} together with
$\bm{\beta}\rightarrow \bm{\beta}^d$, results in
\begin{equation}
\label{UDCnonret}
U_{DC}^{\mathrm{nr}}(\vect{R})=-
\frac{5\mi\mu_0^2c}{64\pi^3 R^6} \epsilon_{ipq} \tilde{R}_q 
\beta^{dij}_{\rma}\left(3\delta_{jr} -7 \tilde{R}_j \tilde{R}_r \right) \sum_{m} \left( \vect{d}_A^{m0}\vect{m}_A^{0m} \right)^{pr}
\,.
\end{equation}

\subsection{Chiral--chiral interaction}
The contribution from the chirality of two molecules A and B
to the vdW interaction potential in free space can be obtained by 
making use of Eqs.~\eqref{curlG} and \eqref{curlGcurl} for the required derivatives of 
the Green tensor in Eq.~\eqref{eq58}, together with Eqs.~\eqref{G0} and \eqref{f&g}.
This leads to
\begin{eqnarray}
\label{CCfree}
&&\hspace{-1in}U_{CC}(\vect{R})
=\frac{\hbar\mu_0^2c^3}{16\pi^3}\int_0^\infty\dif k\,k^6\me^{-2kR}\chi_\rma^{ij}(\mi k c)\chi_\rmb^{pq}(\mi k c)\nonumber\\
&&\hspace{-.9in}\times\bigg\{
\left(\frac{1}{k^6 R^6}+\frac{2}{k^5R^5}\right)
\left(\delta_{jq}-3\tilde{R}_j\tilde{R}_q\right)
\left(\delta_{ip}-3\tilde{R}_i\tilde{R}_p\right)
\nonumber\\
&&\hspace{-.6in}+\frac{1}{k^4 R^4}\left[3\delta_{jq}\delta_{ip}-
7(\delta_{jq}\tilde{R}_i\tilde{R}_p
+\delta_{ip}\tilde{R}_j\tilde{R}_q)
+15\tilde{R}_i\tilde{R}_j\tilde{R}_p\tilde{R}_q
+\epsilon_{jrp}\epsilon_{qsi}\tilde{R}_r\tilde{R}_s\right]\nonumber\\
&&\hspace{-.6in}+\frac{1}{k^3 R^3}\left[2\delta_{jq}\delta_{ip}-
4(\delta_{jq}\tilde{R}_i\tilde{R}_p+\delta_{ip}\tilde{R}_j
\tilde{R}_q)
+6\tilde{R}_i\tilde{R}_j\tilde{R}_p\tilde{R}_q
+2\epsilon_{jrp}\epsilon_{qsi}\tilde{R}_r\tilde{R}_s\right]
\nonumber\\
&&\hspace{-.6in}+\frac{1}{k^2 R^2}\left[\delta_{jq}\delta_{ip}-
(\delta_{jq}\tilde{R}_i\tilde{R}_p+\delta_{ip}\tilde{R}_j
\tilde{R}_q)
+\tilde{R}_i\tilde{R}_j\tilde{R}_p\tilde{R}_q
+\epsilon_{jrp}\epsilon_{qsi}\tilde{R}_r\tilde{R}_s\right]
\bigg\}.
\end{eqnarray}
It is worth noting that in the case of isotropic molecules, for which
$\chi_{\rma'}^{ij} \equiv \chi_{\rma'}\delta_{ij}$, Eq.~\eqref{CCfree} 
reduces to
\begin{eqnarray}
&&\hspace{-1in}U_{CC}(\vect{R})
=\frac{\hbar\mu_0^2c^3}{8\pi^3R^6}\int_0^\infty\dif k\,\me^{-2kR}\chi_\rma(\mi k c)\chi_\rmb(\mi k c)
\left(3+6kR+4k^2R^2\right)
\end{eqnarray}
in agreement with Refs.~\cite{jenkins1994-1, jenkins1994-2, salam1996, craig1999}. As stated above, this interaction cannot be obtained from the well-known electric--electric potential by means of a duality transformation.

In the retarded intermolecular separation, the chiral polarisabilities in
Eq.~(\ref{CCfree}) can be replaced by their static limits. Calculation of the remaining integral results in
\begin{align}
&U^\text{r}_{CC}(\vect{R})
=\frac{\hbar\mu_0^2c^3}{128\pi^3R^9}\chi_\rma^{\prime ij}\chi_\rmb^{\prime pq}\nonumber\\
&\,\,\times
\left(101\delta_{ip}\delta_{jq}
-171\delta_{jq}\tilde{R}_i\tilde{R}_p-
171\delta_{ip}
\tilde{R}_j\tilde{R}_q
+297\tilde{R}_i\tilde{R}_j\tilde{R}_p\tilde{R}_q
+81\epsilon_{jrp}\epsilon_{qsi}\tilde{R}_r\tilde{R}_s
\right),
\end{align}
where $\chi'_{\rma'}$ is defined according to Eq.~\eqref{alpha0}. 

In the opposite limit of non-retarded intermolecular distances, we may set the exponential factor in Eq.~\eqref{CCfree} to unity and retain only the term proportional to $\frac{1}{k^6R^6}$ in the curly brackets of Eq.~(\ref{CCfree}). Doing this together with using Eq.~\eqref{chiem} for chiral polarisabilities  and performing the remaining integral leads to 
\begin{multline}
U_{CC}^{\text{nr}}(\vect{R})=-\frac{\mu_0^2c^2}{8 \pi^2 \hbar R^6} 
\left(\delta_{ip}-3\tilde{R}_i\tilde{R}_p\right)
\left(\delta_{jq}-3\tilde{R}_j\tilde{R}_q\right)
 \sum_{l,m} \frac{\left( \vect{d}_A^{m0}\vect{m}_A^{0m} \right)^{ij} \left( \vect{d}_B^{l0}\vect{m}_B^{0l} \right)^{pq}}{\omega_A^m+\omega_B^l}\,.
\end{multline}

\section{Conclusion}
\label{conclusion}
In this paper a general expression for the van der Waals dispersion interaction potential between two ground-state molecules in the presence of arbitrary magnetoelectric media is derived making use of perturbation theory. 
The result for the diamagnetic--chiral interaction potential is seen to have 
the same form as that of parmagnetic--chiral one, despite being obtained from different 
orders of perturbative calculations. The formulas obtained in this paper, together with 
those found in earlier studies (e.g.~Refs.~\cite{safari2006, buhmann2013-3, safari2008}) form a complete set of formulas for the van der Waals interaction potentials for molecules with electric, paramagnetic, and diamagnetic polarisabilities, applicable for chiral and anisotropic molecules and for arbitrary ranges of intermolecular separations as well as allowing for a possible chirality of the surrounding media.
The formulas of various contributions to the full dispersion interaction 
have been brought into a unified form by taking advantage of electric--magnetic duality. 

As an application of the obtained formulas, the interaction potential for two anisotropic chiral molecules in free space has been evaluated.
The retarded and non-retarded limits of intermolecular separations are obtained and their corresponding distance power laws are summarized in Tab.~\ref{powerlaw}.
\begin{table}[!htbp]
\caption{
\hspace{-3ex} 
 \hspace{1ex}
Distance power laws for the various dispersion potentials. }
\label{powerlaw}
\centering

\begin{tabular}{|c|c|c|}
\hline
Interaction component & Retarded limit & non-retarded limit   \\
\hline
\hline
electric--electric \cite{safari2006}& $-R^{-7}$ & $-R^{-6}$ \\
\hline
electric--paramagnetic \cite{safari2008}& $+R^{-7}$ & $+R^{-4}$ \\
\hline
electric--diamagnetic \cite{buhmann2013-3}& $-R^{-7}$ & $-R^{-4}$ \\
\hline
electric--chiral \cite{SalamThesis} & $\pm R^{-8}$ & $\pm R^{-5}$ \\
\hline
paramagnetic--paramagnetic \cite{safari2008}& $-R^{-7}$ & $-R^{-6}$ \\
\hline
paramagnetic--diamagnetic  \cite{buhmann2013-3}& $+R^{-7}$ & $+R^{-6}$ \\
\hline
paramagnetic--chiral \cite{SalamThesis}& $\pm R^{-8}$ & $\pm R^{-5}$ \\
\hline
diamagnetic--diamagnetic \cite{buhmann2013-3}& $-R^{-7}$ & $-R^{-6}$ \\
\hline
diamagnetic--chiral \cite{SalamUnpublished}& $\pm R^{-8}$ & $\pm R^{-6}$ \\
\hline
chiral--chiral \cite{Barcellona2017}& $\pm R^{-9}$ & $\pm R^{-6}$ \\
\hline
\end{tabular}
\end{table}

\ack
AS and SYB thank J.~Franz for discussions. SYB is grateful for support by the German Research Council (grant
BU 1803/3-1, S.Y.B.) and the Freiburg Institute for Advanced Studies.
AS acknowledges the award of a Mercator Fellowship funded by the Deutsche Forschungsgemeinschaft through the IRTG 2079/Cold Controlled Ensembles in Physics and Chemistry at the University of Freiburg. He also thanks the Freiburg Institute for Advanced Studies (FRIAS), University of Freiburg for the award of a Marie S Curie External Senior Fellowship under the EU Horizon 2020 Grant No. 75430.

\begin{appendix}
\section{Calculation of Eq.~(\ref{eq33}) for the chiral--electric potential}
\label{AppA}
Taking into account the contributions from all diagrams in Fig.~\ref{fig1} to Eq.~(\ref{p4}) for the electric--chiral interaction
leads to 
\begin{eqnarray}
\label{eq30}
&&\hspace{-.9in}U_{EC}=\sum_{i=1}^{12} U^{i}_{EC}\nonumber\\
&&\hspace{-.8in}=-\frac{\mu_0^2}{\hbar\pi^2}
\sum_{m,l\neq 0}\int_0^\infty\dif\omega_1\int_0^\infty\dif\omega_2 \,
\omega_1^2\omega_2^2 N_{EC}(\omega_1,\omega_2)[D^+_{EC}(\omega_1,\omega_2)+D^-_{EC}(\omega_2,\omega_1)],\nonumber\\
\end{eqnarray}    
where 
\begin{align}
\label{eq31}
D^{\pm}_{EC}(\omega_1,\omega_2)=&\pm\left(\frac{1}{D_1}-\frac{1}{D_2}+\frac{1}{D_3}
-\frac{1}{D_9}-\frac{1}{D_{11}}-\frac{1}{D_{12}}\right)+\frac{1}{D_4}-\frac{1}{D_5}+\frac{1}{D_6}\nonumber\\
&
+\frac{1}{D_7}+\frac{1}{D_8}+\frac{1}{D_{10}}\end{align}
with $D_{i}$ denoting the energy denominator for case ($i$) as listed 
in Tab.~\ref{tab1}. It is not difficult to show that 
\begin{align}
\label{eq32}
&D^+_{EC}(\omega_1,\omega_2)+D^-_{EC}(\omega_2,\omega_1)\nonumber\\
&=\frac{4\omega_\rma^m}{\omega_\rma^m+\omega_\rmb^l}\bigg[\frac{1}{(\omega_\rma^m+\omega_2)(\omega_\rmb^l+\omega_2)}-\frac{1}{(\omega_\rma^m+\omega_1)(\omega_\rmb^l+\omega_1)}\bigg]\left(\frac{1}{\omega_2+\omega_1}
-\frac{1}{\omega_2-\omega_1}\right)
\end{align}
(cf.~Ref.~\cite{craig2012}). Making use of Eq.~(\ref{eq32}) in Eq.~(\ref{eq30}) leads to Eq.~(\ref{eq33}).
\begin{table}
\centering\begin{tabular}{ll}
\hline
Diagram  & \hspace{9ex}Energy denominator $D_{i}$\\
\hline
\vspace{1ex}
(1)  &\qquad $D_{1}=(\omega_{\rma}^{m}+\omega_1)(\omega_1+\omega_2)(\omega_{\rmb}^{l}+\omega_2)$  \\
\vspace{1ex}
(2)  & $\qquad D_{2}=(\omega_{\rma}^{m}+\omega_1)(\omega_1+\omega_2)(\omega_{\rmb}^{l}+\omega_1)$\\
\vspace{1ex}
(3)  & $\qquad D_{3}=(\omega_{\rma}^m+\omega_1)(\omega_{\rma}^m+\omega_{\rmb}^l)(\omega_{\rmb}^l+\omega_2)$\\
\vspace{1ex}
(4)  & $\qquad D_{4}=(\omega_{\rma}^m+\omega_1)(\omega_{\rma}^m+\omega_{\rmb}^l+\omega_1+\omega_2)(\omega_{\rmb}^l+\omega_1)$\\
\vspace{1ex}
(5)  & $\qquad D_{5}=(\omega_{\rma}^m+\omega_1)(\omega_{\rma}^m+\omega_{\rmb}^l)(\omega_{\rma}^m+\omega_2)$\\
\vspace{1ex}
(6)  & $\qquad D_{6}=(\omega_{\rma}^m+\omega_1)(\omega_{\rma}^m+\omega_{\rmb}^l+\omega_1+\omega_2)(\omega_{\rma}^m+\omega_2)$\\
\vspace{1ex}
(7)  & $\qquad D_{7}=(\omega_{\rmb}^l+\omega_1)(\omega_{\rma}^m+\omega_{\rmb}^l)(\omega_{\rmb}^l+\omega_2)$\\
\vspace{1ex}
(8)  & $\qquad D_{8}=(\omega_{\rmb}^l+\omega_1)(\omega_{\rma}^m+\omega_{\rmb}^l+\omega_1+\omega_2)(\omega_{\rmb}^l+\omega_2)$\\
\vspace{1ex}
(9)  & $\qquad D_{9}=(\omega_{\rmb}^l+\omega_1)(\omega_{\rma}^m+\omega_{\rmb}^l)(\omega_{\rma}^m+\omega_2)$\\
\vspace{1ex}
(10)  & $\qquad D_{10}=(\omega_{\rmb}^l+\omega_1)(\omega_{\rma}^m+\omega_{\rmb}^l+\omega_1+\omega_2)(\omega_{\rma}^m+\omega_1)$\\
\vspace{1ex}
(11)  & $\qquad D_{11}=(\omega_{\rmb}^l+\omega_1)(\omega_1+\omega_2)(\omega_{\rma}^m+\omega_2)$\\
\vspace{1ex}
(12)  & $\qquad D_{12}=(\omega_{\rmb}^l+\omega_1)(\omega_1+\omega_2)(\omega_{\rma}^m+\omega_1)$\\
\hline
\end{tabular}
\caption{
\label{tab1}
The energy denominators corresponding to the diagrams in Fig.~\ref{fig1}.
}
\end{table}
\section{Calculation of Eq.~(\ref{eq34}) for the chiral--electric potential}
\label{AppB}
In Eq.~(\ref{eq33}) we encounter a double integral in the form of 
\begin{align}
\label{a1}
\int_0^\infty\dif\omega_1
&\int_0^\infty\dif\omega_2\,
\frac{\omega_1^2\omega_2\,\im{G}_\alpha(\omega_1)\im
{G}_\beta(\omega_2)}{(\omega_{\rma}^m+\omega_i)(\omega_{\rmb}^l+\omega_i)}
\bigg(\frac{1}{\omega_2+\omega_1}-\frac{1}{\omega_2-\omega_1}\bigg)\equiv I_i,
 \quad i=1,2
\end{align}
where $G_\alpha$ and $G_\beta$ denote typical matrix elements of the Green tensor. 
In order to calculate $I_1$, the integral over $\omega_2$,
\begin{equation}
\label{a2}
\int_0^\infty d\omega_2 \omega_2
\im G_\beta(\omega_2)\left(\frac{1}{\omega_2+\omega_1}-\frac{1}{\omega_2-\omega_1}\right)
\equiv g_\beta^{(1)}(\omega_1)
\end{equation}
is to be performed first, for which we introduce its more general form as
\begin{equation}
\label{a3}
\bm{g}^{(n)} (\omega)= \int_0^\infty\dif\omega'
\,{\omega'}^{n}\im \tens{G}(\omega')
\bigg(\frac{1}{\omega'+\omega}+\frac{(-1)^n}{\omega'-\omega}\bigg),
\end{equation}
for real $\omega$ and non-negative integer $n$. The Schwartz reflection principle, Eq.~(\ref{Schwartz}), 
implies that for real $\omega$ the imaginary part of the Green tensor is an odd function of $\omega$. 
Using this, it is easy to show that the integrand in Eq.~(\ref{a3}) is
an even function of $\omega'$, and we may write
\begin{equation}
\label{a4}
\bm{g}^{(n)} (\omega)=\frac{1}{2}\im \int_{-\infty}^\infty\dif\omega'
\,{\omega'}^{n} \tens{G}(\omega')
\bigg(\frac{1}{\omega'+\omega}+\frac{(-1)^n}{\omega'-\omega}\bigg). 
\end{equation}
This integral can be evaluated using contour integral techniques by drawing an infinitely large semicircle
in the upper half of the complex frequency plane (where the Green's function is analytic as a response function \cite{dung2003}) to the real axis, and making use of the Cauchy formula. This leads to
 \begin{equation}
\label{a5}
\bm{g}^{(n)} (\omega)= \frac{\pi}{2}(-\omega)^{n}[\tens{G}(\omega)+\tens{G}^\ast(\omega)].
\end{equation}
Using Eq.~(\ref{a2}) together with Eq.~(\ref{a5}) in Eq.~(\ref{a1}) results in
\begin{eqnarray}
\label{a6}
&&\hspace{-.9in}I_1=\frac{\pi \mi}{4}\int_0^\infty
\frac{\dif\omega\,\omega^3}{(\omega_{\rma}^m+\omega)(\omega_{\rmb}^l+\omega)} 
\left[G_\alpha(\omega)-G^\ast_\alpha(\omega)\right]
\left[G_\beta(\omega)+G^\ast_\beta(\omega)\right]\nonumber\\
&&\hspace{-.7in}=\frac{\pi \mi}{4}\int_0^\infty
\frac{\dif\omega\,\omega^3}{(\omega_{\rma}^m+\omega)(\omega_{\rmb}^l+\omega)} 
\left[G_\alpha(\omega)G_\beta(\omega)-G^\ast_\alpha(\omega)G^\ast_\beta(\omega)\right]\nonumber\\
&&\hspace{-.5in}+\frac{\pi \mi}{4}\int_0^\infty
\frac{\dif\omega\,\omega^3}{(\omega_{\rma}^m+\omega)(\omega_{\rmb}^l+\omega)} 
\left[G_\alpha(\omega)G^\ast_\beta(\omega)-G^\ast_\alpha(\omega)G_\beta(\omega)\right].
\end{eqnarray}
The double integral $I_2$ defined by Eq.~(\ref{a1}) can be treated similarly to obtain
\begin{eqnarray}
\label{a7}
&&\hspace{-.9in}I_2=-\frac{\pi \mi}{4}\int_0^\infty
\frac{\dif\omega\,\omega^3}{(\omega_{\rma}^m+\omega)(\omega_{\rmb}^l+\omega)} 
\left[G_\alpha(\omega)G_\beta(\omega)-G^\ast_\alpha(\omega)G^\ast_\beta(\omega)\right]\nonumber\\
&&\hspace{-.6in}+\frac{\pi \mi}{4}\int_0^\infty
\frac{\dif\omega\,\omega^3}{(\omega_{\rma}^m+\omega)(\omega_{\rmb}^l+\omega)} 
\left[G_\alpha(\omega)G^\ast_\beta(\omega)-G^\ast_\alpha(\omega)G_\beta(\omega)\right],
\end{eqnarray}
hence, for $I_2-I_1$ appearing in Eq.~(\ref{eq33}), we end up with
\begin{equation}
\label{a8}
I_2-I_1=-\frac{\pi \mi}{2}\int_0^\infty
\frac{\dif\omega\,\omega^3}{(\omega_{\rma}^m+\omega)(\omega_{\rmb}^l+\omega)} 
\left[G_\alpha(\omega)G_\beta(\omega)-G^\ast_\alpha(\omega)G^\ast_\beta(\omega)\right].
\end{equation}    
For the second term in the square brackets we may change the variable as $\omega\to-\omega$ and make use of the Schwartz reflection principle to rewrite Eq.~(\ref{a8}) in the form
\begin{equation}
\label{a9}
I_2-I_1=-\frac{\pi \mi}{2}
\left\{\int_0^\infty 
 \frac{\dif\omega\,\omega^3 G_\alpha(\omega)G_\beta(\omega)}
{(\omega_{\rma}^m+\omega)(\omega_{\rmb}^l+\omega)}+
\int_{-\infty}^0 
 \frac{\dif\omega\,\omega^3 G_\alpha(\omega)G_\beta(\omega)}
 {(\omega_{\rma}^m-\omega)(\omega_{\rmb}^l-\omega)}
 \right\}.
\end{equation} 
Again, using the fact that both of the integrands are analytic in the upper half of the complex frequency plane   including the real axis, we may replace the integration path in the first(second) integral by
a quarter-circle in the first(second) quadrant together with the positive imaginary axis ($\omega$ $\!=\mi \xi$).
The contribution from the quarter circle vanishes due to the limiting behaviour of the Green tensor for large frequencies, and we are left with  
\begin{equation}
I_2-I_1  = -\pi(\omega_{\rma}^m+\omega_{\rmb}^l)\int_0^\infty \dif \xi\frac{\xi^4
{G}_\alpha(\mi \xi){G}_\beta(\mi \xi)}{[(\omega_{\rma}^m)^2+\xi^2][(\omega_{\rmb}^l)^2+\xi^2]}.
\end{equation}
Using this, leads from Eq.~(\ref{eq33}) to Eq.~(\ref{eq34}).

\section{Calculation of Eq.~(\ref{eq44}) for the paramagnetic--chiral potential}
\label{AppC}
\begin{eqnarray}
\label{eq42}
&&\hspace{-1in}U_{PC}(\rr_\rma,\rr_\rmb)\nonumber\\
&&\hspace{-1in}=-\frac{\mu_0^2}{\hbar\pi^2}
\sum_{m,l\neq 0}\int_0^\infty\dif\omega_1\int_0^\infty\dif\omega_2 \,
\omega_1^2\omega_2^2 N_{PC}(\omega_1,\omega_2)[D^+_{PC}(\omega_1,\omega_2)+D^-_{PC}(\omega_2,\omega_1)],
\end{eqnarray}    
where
\begin{align}
\label{eq43}
D^{\pm}_{PC}(\omega_1,\omega_2)=&\pm\left(\frac{1}{D_1}-\frac{1}{D_2}+\frac{1}{D_3}-\frac{1}{D_9}-\frac{1}{D_{11}}
-\frac{1}{D_{12}}\right)-\frac{1}{D_4}+\frac{1}{D_5}-\frac{1}{D_{6}}\nonumber\\
&-\frac{1}{D_7}-\frac{1}{D_8}-\frac{1}{D_{10}}
\,.
\end{align}
By a straightforward algebra it can be shown that 
\begin{align}
\label{eq43-2}&D^+_{PC}(\omega_1,\omega_2)+D^-_{PC}(\omega_2,\omega_1)\nonumber\\
&=\frac{4\omega_\rma^m}{\omega_\rma^m+\omega_\rmb^l}\bigg[\frac{1}{(\omega_\rma^m+\omega_2)(\omega_\rmb^l+\omega_2)}-\frac{1}{(\omega_\rma^m+\omega_1)(\omega_\rmb^l+\omega_1)}\bigg]\left(\frac{1}{\omega_2+\omega_1}
+\frac{1}{\omega_2-\omega_1}\right).
\end{align}
Substituting Eqs.~(\ref{eq41}) and (\ref{eq43-2}) into Eq.~(\ref{eq42}) leads to Eq.~(\ref{eq44}).
\section{Calculation of Eq.~(\ref{eq58}) for the chiral--chiral potential}
\label{AppD}
By calculating the contribution of all other diagrams in Fig.~\ref{fig1} similar to that of diagram (1) 
and summing them up, we obtain
\begin{align}
\label{eq53}
U_{CC}=-
\frac{\mu_0^2}{\hbar\pi^2}\sum_{m,l\neq 0} \int_0^\infty\dif\omega_1
\int_0^\infty&\dif\omega_2\,{\omega_1^2\omega_2^2}
\left\{N^{1a}_{CC}(\omega_1,\omega_2)[{D}^-(\omega_1,\omega_2)+{D}^-(\omega_2,\omega_1)]\right.\nonumber\\
&
\left.
+N^{1b}_{CC}(\omega_1,\omega_2)[{D}^+(\omega_1,\omega_2)+D^+(\omega_2,\omega_1)]\right\}
\end{align}
with \begin{align}
{D}^{\pm}(\omega_1,\omega_2)=&\frac{1}{D_1}-\frac{1}{D_2}+\frac{1}{D_3}+\frac{1}{D_9}+\frac{1}{D_{11}}-\frac{1}{D_{12}}
\nonumber\\
&
\pm\left(\frac{1}{D_4}-\frac{1}{D_5}+\frac{1}{D_6}-\frac{1}{D_7}+\frac{1}{D_8}+\frac{1}{D_{10}}\right).
\end{align}
After some straightforward algebra, it can be shown that
\begin{equation}
{D}^{\pm}(\omega_1,\omega_2)+{D}^{\pm}(\omega_2,\omega_1)=\frac{4}{\omega_\rma^m+\omega_\rmb^l}F_{\pm}(\omega_1,\omega_2),
\end{equation}
 where $F_{\pm}(\omega_1,\omega_2)=f_{\pm}(\omega_1,\omega_2)+f_{\pm}(\omega_2,\omega_1)$ with
 \begin{equation}
f_{\pm}(\omega_1,\omega_2)=\frac{\omega_1}{(\omega_\rma^m+\omega_1)(\omega_\rmb^l+\omega_1)}\bigg(\frac{1}{\omega_1+\omega_2}
\pm\frac{1}{\omega_1-\omega_2}\bigg).
\end{equation}
Upon making use of Eqs.~(\ref{eq49}) and (\ref{eq50}), Eq.~(\ref{eq53}) becomes
\begin{eqnarray}
&&\hspace{-1.in}U_{CC}=
-\frac{4\mu_0^2}{\hbar\pi^2}\sum_{m,l\neq 
0}\frac{1}{\omega_\rma^m+\omega_\rmb^l} \int_0^\infty\dif\omega_1
\int_0^\infty\dif\omega_2\,\omega_1
\nonumber\\
&&\hspace{-1.in}\times\bigg\{\omega_1\trace\left[
\vect{d}_\rma^{m0}\vect
{m}_{\rma }^{0m} \cdot\dela\times\im
\tens{G}(\vect{r}_\rma,\vect{r}_\rmb,\omega_2)\times\overleftarrow{\nabla}
_\rmb\cdot 
\vect{m}_{\rmb}^{l0}\vect{d}_\rmb^{0l}\cdot\im\tens{G}(\vect{r}_\rmb,\vect{r}_\rma,
\omega_1)\right]F_-
\nonumber\\
&&\hspace{-.8in}-\omega_2\trace\left[\vect{d}_{\rma}^{m0}\vect{m}_{\rma
} ^{0m}
\cdot\dela\times\im\tens{G
}(\vect{r}_\rma,\vect{r}_\rmb,\omega_2)
\cdot\vect{d}_\rmb^{l0}\vect{m}_{\rmb}^{0l}\cdot\delb\times\im\tens{G}(\vect
{
r}_\rmb,\vect{r}_\rma,\omega_1)
\right]F_+\bigg\}.\nonumber\\
\end{eqnarray}
Now, in order to write the result in a compact form, we follow a similar procedure as for obtaining Eqs.~(\ref{eq35}) and (\ref{eq45}).
\section{Calculation of Eq.~(\ref{eq68}) for the diamagnetic--chiral potential}
\label{AppE}
In order to determine the diamagnetic--chiral interaction potential,
Eq.~(\ref{eq65}) has to be added to other portions for those 
$D_{1(a)}$ is replaced by $D_{ia(b)}$ given by Eq.~(\ref{eq67}). Doing this yields    
\begin{align}
&U_{DC}= -\frac{4\mu_0^2\mi}{\pi^2}\sum_l
\int_0^\infty\dif\omega_1\int_0^\infty\dif\omega_2 \frac{\omega_1\omega_2}{(\omega_{\rmb}+\omega_1)(\omega_1+\omega_2)(\omega_{\rmb}+\omega_2)}
\nonumber\\
&\times\trace\big\{
\bm{\beta}_\rma\cdot\big[\dela\times\im\tens{G}(\vect{r}_{\rma},\vect{r}_{\rmb},\omega_2)\times\ledb\big]\cdot\vect{m}^{l0}_{\rmb}\vect{d}^{0l}_{\rmb}
\cdot\big[\im\tens{G}(\vect{r}_{\rmb},\vect{r}_{\rma},\omega_1)\times\leda\big]
\big\}.
\end{align}
This may be written as
\begin{align}
\label{B1}
U_{DC}=-\frac{4\mu_0^2\mi}{\pi^2}&\sum_l 
\int_0^\infty\dif\omega_1 \frac{\omega_1}{(\omega_{\rmb}^l+\omega_1)}
\nonumber\\
&\times\trace\big\{
\bm{\beta}_\rma\cdot\dela\times \tens{J}_1\times\ledb\cdot\vect{m}^{l0}_{\rmb}\vect{d}^{0l}_{\rmb}
\cdot\big[\im\tens{G}(\vect{r}_{\rmb},\vect{r}_{\rma},\omega_1)\times\leda\big]\big\}
\end{align}
with
\begin{eqnarray}
\label{B2}
&&\tens{J}_1(\omega)=\int_0^\infty\dif\omega_2 
\frac{\omega_2}{(\omega+\omega_2)(\omega_{\rmb}^l+\omega_2)}
\im\tens{G}(\vect{r}_{\rma},\vect{r}_{\rmb},\omega_2).
\end{eqnarray}
The Green's tensor being analytic in the upper half of the complex frequency plane, facilitates rewriting $\vect{J}_1(\omega)$ in terms of the imaginary frequency $\omega_2\to\mi \xi$ as
\begin{align}
\label{B3}
\tens{J}_1(\omega)&=\im\int_0^\infty\dif\omega_2 \frac{\omega_2}{(\omega+\omega_2)(\omega_{\rmb}^l+\omega_2)}
\tens{G}(\vect{r}_{\rma},\vect{r}_{\rmb},\omega_2)&\nonumber\\
&=
\int_0^\infty\dif \xi \xi^2\frac{(\omega+\omega_{\rmb}^l)}{(\omega^2+\xi^2)[(\omega_{\rmb}^l)^2+\xi^2]}\tens{G}(\vect{r}_{\rma},\vect{r}_{\rmb},\mi \xi).
\end{align}
In the second line we have used the fact that the Green tensor is real-valued for imaginary frequency due to the Schwartz reflection principle, Eq.~(\ref{Schwartz}). Using Eq.~(\ref{B3}) in (\ref{B1}) gives
 \begin{align}
\label{B4}
U_{DC} = -\frac{4\mu_0^2\mi}{\pi^2}\sum_l &\int_0^\infty\dif \xi \xi^2\frac{1}{[(\omega_{\rmb}^l)^2+\xi^2]}\nonumber\\
&\times
\trace\big\{
\bm{\beta}_\rma\cdot\dela\times\tens{G}(\vect{r}_{\rma},\vect{r}_{\rmb},\mi \xi) \times\ledb\cdot\vect{m}^{l0}_{\rmb}\vect{d}^{0l}_{\rmb}
\cdot\big[\tens{J}_2\times\leda\big]\big\}
\end{align}
with 
\begin{eqnarray}
\label{B5}
&&\tens{J}_2=\im\int_0^\infty\dif\omega\frac{\omega}{(\omega^2+\xi^2)} 
\tens{G}(\vect{r}_{\rmb},\vect{r}_{\rma},\omega).
\end{eqnarray}
At this stage, noting that the integrand in (\ref{B5})
has a simple pole at $\omega=\mi \xi$ in the upper half of the complex frequency plane, we may use contour integral
 techniques to replace the integration path with the positive part of the imaginary axis and excluding the pole by applying an infinitesimal halfcircle around it. 
 Doing so, we find
\begin{align}
\label{B6}
\int_0^\infty\dif\omega&\frac{\omega}{(\omega^2+\xi^2)} 
\tens{G}(\vect{r}_{\rmb},\vect{r}_{\rma},\omega) \nonumber\\
&=-P\int_0^\infty\dif v\frac{v}{(-v^2+\xi^2)} 
\tens{G}(\vect{r}_{\rmb},\vect{r}_{\rma},\mi v)
+\mi  \frac{\pi }{2}\tens{G}(\vect{r}_{\rmb},\vect{r}_{\rma},\mi \xi).
\end{align}
Hence, the right hand side of Eq.~(\ref{B5}) reduces to 
$(\pi/2)\tens{G}(\vect{r}_{\rmb},\vect{r}_{\rma},\mi \xi)$. 
Using this in Eq.~(\ref{B4}) results in
\begin{align}
\label{B7}
U_{DC} =& -\frac{2\mu_0^2\mi}{\pi}\sum_l
\int_0^\infty\dif \xi\frac{\xi^2}{(\omega_{\rmb}^l)^2+\xi^2}\nonumber\\
&\times
\trace\big\{
\bm{\beta}_\rma\cdot\dela\times\tens{G}(\vect{r}_{\rma},\vect{r}_{\rmb},\mi \xi) \times\ledb\cdot\vect{m}^{l0}_{\rmb}\vect{d}^{0l}_{\rmb}
\cdot\big[\tens{G}(\vect{r}_{\rmb},\vect{r}_{\rma},\mi \xi)\times\leda\big]\big\}.
\end{align}
Finally, the summation over $l$ in the right hand side can be replaced in terms of the chiral polarisability defined by Eq.~(\ref{chime}),
 \begin{equation}
\sum_l\frac{\vect{m}^{l0}_{\rmb}\vect{d}^{0l}_{\rmb}}{(\omega_{\rmb}^l)^2+\xi^2}=-\frac{\hbar}{2\mi \xi}\bm{\chi}_{\rmb}^{me}(\mi \xi),
\end{equation}
which leads to Eq.~(\ref{eq68}). 
 
 \end{appendix}

\section*{References}
\bibliography{Refs}

\end{document}